\documentclass[12pt]{myiopart}
\usepackage{iopams}
\usepackage{bm}
\usepackage{graphicx}
\usepackage{cite}
\usepackage{multirow}
\usepackage{mathtools}
\usepackage{url}
\usepackage[hyperindex]{hyperref}
\usepackage{subfigure}
\usepackage{stackrel}
\usepackage{xspace}
\newcommand{\nua}[1]{\ensuremath{\rlap{\kern-2.5pt\ensuremath{\overset{\scriptscriptstyle(-)}{\phantom{\nu}}}}{\ensuremath{{\nu}_{#1}}}}\xspace}
\newcommand{\openone}{\ensuremath{\bm{1}}\xspace}
\newcommand{\eff}{\ensuremath{\mathrm{eff}}\xspace}

\newcommand{\meff}[1]{\ensuremath{m^{\eff}_{#1}}\xspace}
\newcommand{\Neff}{\ensuremath{N_{\eff}}\xspace}
\newcommand{\mnu}{\ensuremath{\sum m_\nu}\xspace}
\newcommand{\DNeff}{\ensuremath{\Delta N_{\eff}}\xspace}

\newcommand{\zeq}{\ensuremath{z_\mathrm{eq}}\xspace}

\newcommand{\lcdm}{$\Lambda$CDM\xspace}
\newcommand{\Hou}{\ensuremath{\, \text{Km s}^{-1}\text{ Mpc}^{-1}}\xspace}
\begin{document}

\topical[Light sterile neutrinos]{Light sterile neutrinos\footnote{
This review is dedicated to the memory of Hai-Wei Long,
our dear friend and collaborator,
who passed away on 29 May 2015.
He was an exceptionally kind person and an enthusiastic physicist.
We deeply miss him.
}}

\author{S~Gariazzo$^{1,2}$, C~Giunti$^{2}$, M~Laveder$^{3,4}$, Y~F~Li$^5$, E~M~Zavanin$^{1,2,6}$}

\address{$^1$ Department of Physics, University of Torino, Via P. Giuria 1, I--10125 Torino, Italy}
\address{$^2$ INFN, Sezione di Torino, Via P. Giuria 1, I--10125 Torino, Italy}
\address{$^3$ Dipartimento di Fisica e Astronomia ``G. Galilei'', Universit\`a di Padova, Italy}
\address{$^4$ INFN, Sezione di Padova, Via F. Marzolo 8, I--35131 Padova, Italy}
\address{$^5$ Institute of High Energy Physics, Chinese Academy of Sciences, Beijing 100049, China}
\address{$^6$ Instituto de F\'isica Gleb Wataghin, Universidade Estadual de Campinas - UNICAMP,
              Rua S\'ergio Buarque de Holanda 777, 13083-859 Campinas SP Brazil}

\ead{gariazzo@to.infn.it,
giunti@to.infn.it,
laveder@pd.infn.it,
liyufeng@ihep.ac.cn,
zavanin@gmail.com}

\begin{abstract}
The theory and phenomenology of
light sterile neutrinos at the eV mass scale is reviewed.
The reactor, Gallium and LSND anomalies are briefly described
and interpreted as indications of the existence of short-baseline oscillations
which require the existence of light sterile neutrinos.
The global fits of short-baseline oscillation data in 3+1 and 3+2 schemes
are discussed,
together with the implications
for $\beta$-decay and neutrinoless double-$\beta$ decay.
The cosmological effects of light sterile neutrinos
are briefly reviewed
and the implications of existing cosmological data
are discussed.
The review concludes with a summary of future perspectives.
\end{abstract}

\pacs{14.60.Pq, 14.60.Lm, 14.60.St, 98.80.-k}




\section{Introduction}
\label{sec:intro}

The possible existence of sterile neutrinos
\cite{Pontecorvo:1968fh},
which do not have the standard weak interactions of the three active neutrinos
$\nu_{e}$,
$\nu_{\mu}$, and
$\nu_{\tau}$,
is currently a hot topic of theoretical and experimental research
which could provide valuable information on the physics beyond the Standard Model
(see
Refs.~\cite{Bilenky:1998dt,Volkas:2001zb,GonzalezGarcia:2002dz,Maltoni:2004ei,Strumia:2006db,GonzalezGarcia:2007ib,Boyarsky:2009ix,Kusenko:2009up,Hannestad:2010kz,Rodejohann:2011mu,Abazajian:2012ys,Rodejohann:2012xd,Conrad:2012qt,Palazzo:2013me,Archidiacono:2013fha,Bellini:2013wra,Drewes:2013gca,Boyarsky:2012rt,1503.01059}).

Sterile neutrinos are singlets of the Standard Model gauge symmetries
which can couple to the active neutrinos through the Lagrangian mass term.
In practice there are bounds on the active-sterile mixing,
but there is no bound on the number of sterile neutrinos and on their mass scales.
Therefore the existence of sterile neutrinos is investigated at different mass scales.
This review is devoted to the discussion of sterile neutrinos at the eV scale,
which can explain the anomalies found in some short-baseline neutrino oscillation experiments
(discussed in details in Section~\ref{sec:sbl}).
However, it is important to remind that there are other very interesting possibilities which are
under study:
very light sterile neutrinos at a mass scale smaller than 0.1 eV, which could affect the oscillations of solar
\cite{deHolanda:2003tx,deHolanda:2010am,Das:2009kw}
and reactor
\cite{Kang:2013zma,Bakhti:2013ora,Palazzo:2013bsa,Girardi:2014wea,An:2014bik}
neutrinos;
sterile neutrinos at the keV scale,
which could constitute warm dark matter according to the Neutrino Minimal Standard Model ($\nu$MSM)
\cite{Asaka:2005an,Asaka:2005pn,Asaka:2006ek,Asaka:2006rw,Asaka:2006nq}
(see also the reviews in Refs.~\cite{Boyarsky:2009ix,Kusenko:2009up,Drewes:2013gca,Boyarsky:2012rt});
sterile neutrinos at the electroweak scale
\cite{Antusch:2015mia,Deppisch:2015qwa}
or above it
\cite{Antusch:2014woa,Deppisch:2015qwa},
whose effects may be seen at LHC and other high-energy colliders.
Let us also note that there are several interesting models
with sterile neutrinos at different mass scales
\cite{Ghosh:2010hy,Barger:2010iv,Grossman:2010iq,deGouvea:2011zz,Chen:2011ai,Barry:2011fp,Fong:2011xh,Zhang:2011vh,Patra:2012ur,Nemevsek:2012iq,Dev:2012bd,Duerr:2013opa,Rodejohann:2014eka,Rosner:2014cha,Adhikari:2014nea,deGouvea:2015pea}.

The plan of this review is as follows.
In Section~\ref{sec:mixing} we introduce the main theoretical aspects of
neutrino mixing and oscillations in the extensions of
standard three-neutrino ($3\nu$) mixing
with light sterile neutrinos.
In Section~\ref{sec:sbl} we summarize the three experimental indications
in favor of short-baseline neutrino oscillations
which require the existence of sterile neutrinos:
the reactor, Gallium and LSND anomalies.
In Section~\ref{sec:global} we discuss the global fits
of short-baseline neutrino oscillation data
in the framework of the simplest mixing schemes with one or two sterile neutrinos at the eV scale.
In Sections~\ref{sec:decay}
and
\ref{sec:cosmo} we review the effects of eV-scale massive neutrinos
in $\beta$ decay and neutrinoless double-$\beta$ decay
and in cosmology.
Finally, in Section~\ref{sec:conclusions}
we present our conclusions and our view of the perspectives
of the search for eV-scale sterile neutrinos.
\section{Neutrino mixing and short-baseline oscillations}
\label{sec:mixing}

The Standard Model of electroweak interactions
\cite{Glashow:1961tr,Weinberg:1967tq,Salam:1968rm}
based on the $\text{SU}(2)_{L} \times \text{U}(1)_{Y}$ gauge symmetry
is a superb theory which can explain the majority of terrestrial experimental observations.
However,
it does not account for neutrino masses,
whose existence have been proved without doubt
by the measurement of neutrino oscillations
in solar, atmospheric and long-baseline
neutrino oscillation experiments
(see
Refs.~\cite{Giunti:2007ry,Bilenky:2010zza,Xing:2011zza,GonzalezGarcia:2012sz,Bellini:2013wra,PDG-2012}).
The simplest way to extend the Standard Model
in order to take into account neutrino masses is through the introduction of
$\text{SU}(2)_{L} \times \text{U}(1)_{Y}$
singlet fields
which are traditionally called ``right-handed neutrino'' fields
or
``sterile neutrino'' fields.
The adjective ``right-handed'' indicates that they do not belong to
$\text{SU}(2)_{L}$
left-handed multiplets.
Therefore,
they are ``sterile'',
because they do not have Standard Model weak interactions.
Moreover,
assuming that they have zero hypercharge\footnote{
We do not consider here the exotic possibility of a small nonzero
hypercharge of the right-handed neutrino fields,
which would imply that neutrinos are Dirac and millicharged particles
(see Refs.~\cite{Foot:1992ui,Giunti:2014ixa}).
},
they are neutral and can be called ``neutrino'' fields.
Many models which extend the Standard Model
include these right-handed sterile neutrino fields
(see Refs.~\cite{Volkas:2001zb,Mohapatra:2004,Mohapatra:2006gs,Boyarsky:2009ix,Abazajian:2012ys,Drewes:2013gca}).
In the following we consider the general theory of neutrino mixing
in which we have the three standard active left-handed flavor neutrino fields
$\nu_{e L}$,
$\nu_{\mu L}$,
$\nu_{\tau L}$
and
$N_{s}$
sterile right-handed flavor neutrino fields
$\nu_{s_{1}R}$,
\ldots,
$\nu_{N_{s}R}$.
The most general Lagrangian mass term which can be written with these fields is
(the superscript ``(F)'' indicates the flavor basis)
\begin{equation}
\mathcal{L}_{\text{mass}}
=
\frac{1}{2} \, {\nu^{(\text{F})}_{L}}^{T} \, \mathcal{C}^{\dagger} \, M \, \nu^{(\text{F})}_{L}
+
\text{H.c.}
,
\label{201}
\end{equation}
where
(the superscripts ``(a)'' and ``(s)'' indicate, respectively, the column matrices of active and sterile neutrino fields)
\begin{equation}
\nu^{(\text{F})}_{L}
=
\begin{pmatrix}
\nu^{(\text{a})}_{L}
\\ \displaystyle
{\nu^{(\text{s})}_{R}}^{c}
\end{pmatrix}
,
\qquad
\nu^{(\text{a})}_{L}
=
\begin{pmatrix}
\nu_{e L}
\\ \displaystyle
\nu_{\mu L}
\\ \displaystyle
\nu_{\tau L}
\end{pmatrix}
,
\qquad
{\nu^{(\text{s})}_{R}}^{c}
=
\begin{pmatrix}
\nu_{s_{1}R}^{c}
\\ \displaystyle
\vdots
\\ \displaystyle
\nu_{s_{N_{s}}R}^{c}
\end{pmatrix}
,
\label{202}
\end{equation}
and
$\mathcal{C}$
is the unitary charge-conjugation matrix\footnote{
We use the notations and conventions in Ref.~\cite{Giunti:2007ry}.
},
such that
$
\mathcal{C}
\,
\gamma_{\mu}^{T}
\,
\mathcal{C}^{-1}
=
- \gamma_{\mu}
$
and
$
\mathcal{C}^{T} = - \mathcal{C}
$.
For any field $\psi$ the charge-conjugated field
$\psi^{c}$
is given by
$\psi^{c} = \mathcal{C} \overline{\psi}^{T}$
and charge conjugation transforms the chirality of a field
(e.g. $\psi_{R}^{c}$ is left-handed).
In general,
$M$ is a complex symmetric mass matrix,
which can be diagonalized with the unitary transformation
(the superscript ``(M)'' indicates the mass basis)
\begin{equation}
\nu^{(\text{F})}_{L}
=
\mathcal{U} \, \nu^{(\text{M})}_{L}
,
\qquad
\text{with}
\qquad
\nu^{(\text{M})}_{L}
=
\begin{pmatrix}
\nu_{1L}
\\ \displaystyle
\vdots
\\ \displaystyle
\nu_{NL}
\end{pmatrix}
,
\label{203}
\end{equation}
where
$N=3+N_{s}$
is the total number of neutrino fields.
The matrix $\mathcal{U}$ is a
$N \times N$
unitary matrix such that
\begin{equation}
\mathcal{U}^{T} M \mathcal{U}
=
\operatorname{diag}\!\left( m_{1}, \ldots, m_{N} \right)
,
\label{204}
\end{equation}
with real and positive masses
$m_{1}, \ldots, m_{N}$
(see Refs.~\cite{Bilenky:1987ty,Giunti:2007ry}).
The Lagrangian mass term (\ref{201}) becomes
\begin{equation}
\mathcal{L}_{\text{mass}}
=
\frac{1}{2}
\sum_{k=1}^{N}
m_{k}
\nu_{kL}^{T} \mathcal{C}^{\dagger} \nu_{kL}
+
\text{H.c.}
=
-
\frac{1}{2}
\sum_{k=1}^{N}
m_{k}
\overline{\nu_{kL}^{c}} \nu_{kL}
+
\text{H.c.}
=
-
\frac{1}{2}
\sum_{k=1}^{N}
m_{k}
\overline{\nu_{k}} \nu_{k}
,
\label{205}
\end{equation}
with the massive Majorana neutrino fields
$\nu_{k} = \nu_{kL} + \nu_{kL}^{c}$
which satisfy the Majorana constraint
$\nu_{k} = \nu_{k}^{c}$.
Hence, in the general case of active-sterile neutrino mixing
the massive neutrinos are Majorana particles\footnote{
However,
it is not excluded that the mixing is such that there are pairs of Majorana neutrino fields with exactly the same mass
which form Dirac neutrino fields.
}.

The physical effects of the unitary transformation (\ref{203})
are due to the non-invariance of the weak interaction Lagrangian.
Let us first consider the leptonic charged-current weak interaction Lagrangian.
In the flavor basis where the mass matrix of the charged leptons
$\ell_{e} \equiv e$,
$\ell_{\mu} \equiv \mu$,
$\ell_{\tau} \equiv \tau$,
is diagonal,
we have
\begin{equation}
\mathcal{L}_{\text{CC}}
=
-
\frac{ g }{ \sqrt{2} }
\sum_{\alpha=e,\mu,\tau}
\overline{\ell_{\alpha L}} \gamma^{\rho} \nu_{\alpha L} W_{\rho}^{\dagger}
+
\text{H.c.}
=
-
\frac{ g }{ \sqrt{2} }
\sum_{k=1}^{N}
\sum_{\alpha=e,\mu,\tau}
\overline{\ell_{\alpha L}} \gamma^{\rho} \mathcal{U}_{\alpha k} \nu_{kL} W_{\rho}^{\dagger}
+
\text{H.c.}
.
\label{f457}
\end{equation}
It is convenient to write $\mathcal{L}_{\text{CC}}$ in the following matrix form
\begin{equation}
\mathcal{L}_{\text{CC}}
=
-
\frac{ g }{ \sqrt{2} }
\overline{\ell_{L}} \gamma^{\rho} \nu^{(\text{a})}_{L} W_{\rho}^{\dagger}
+
\text{H.c.}
=
-
\frac{ g }{ \sqrt{2} }
\overline{\ell_{L}} \gamma^{\rho} U \nu^{(\text{M})}_{L} W_{\rho}^{\dagger}
+
\text{H.c.}
,
\label{206}
\end{equation}
with
\begin{equation}
\ell_{L}
=
\begin{pmatrix}
e
\\
\mu
\\
\tau
\end{pmatrix}
,
\qquad
\nu^{(\text{a})}_{L}
=
U \nu^{(\text{M})}_{L}
\qquad
\text{and}
\qquad
U
=
\left.
\mathcal{U}
\right|_{3 \times N}
.
\label{207}
\end{equation}
The mixing matrix $U$ is a $3 \times N$
rectangular matrix formed by the first three rows of $\mathcal{U}$.
Therefore, the number of physical mixing parameters is smaller than the number necessary to parameterize
the unitary matrix $\mathcal{U}$.
This is due to the arbitrariness of the mixing in the sterile sector,
which does not affect weak interactions.
A careful analysis
(see Ref.~\cite{Giunti:2007ry})
shows that
the $3 \times N$ mixing matrix $U$ can be parameterized
in terms of
$3 + 3 N_{s}$ mixing angles
and
$3 + 3 N_{s}$ physical phases,
of which
$1 + 2 N_{s}$ are Dirac phases
and
$N - 1$ are Majorana phases.
For such parameterization,
it is convenient to use the scheme
\begin{equation}
U
=
\left[
\left(
\prod_{a=1}^{3}
\prod_{b=4}^{N}
W^{ab}
\right)
R^{23}
W^{13}
R^{12}
\right]_{3 \times N}
\operatorname{diag}\!\left(
1, e^{i\lambda_{21}}, \ldots, e^{i\lambda_{N1}}
\right)
.
\label{f470}
\end{equation}
The unitary $N \times N$ matrix
$W^{ab} = W^{ab}(\theta_{ab},\eta_{ab})$
represents a complex rotation in the $a$-$b$ plane
by a mixing angle $\theta_{ab}$ and a Dirac phase $\eta_{ab}$.
Its components are
\begin{equation}
\left[
W^{ab}(\vartheta_{ab},\eta_{ab})
\right]_{rs}
=
\delta_{rs}
+
\left( c_{ab} - 1 \right)
\left(
\delta_{ra} \delta_{sa}
+
\delta_{rb} \delta_{sb}
\right)
+
s_{ab}
\left(
e^{i\eta_{ab}} \delta_{ra} \delta_{sb}
-
e^{-i\eta_{ab}} \delta_{rb} \delta_{sa}
\right)
,
\label{d048}
\end{equation}
where
$c_{ab}\equiv\cos\vartheta_{ab}$
and
$s_{ab}\equiv\sin\vartheta_{ab}$.
The order of the product of $W^{ab}$ matrices in Eq.~(\ref{f470}) is arbitrary.
The orthogonal matrix
$R^{ab}=W^{ab}(\theta_{ab},0)$
represents a real rotation in the $a$-$b$ plane.
The square brackets with subscript $3 \times N$
indicate that the enclosed $N \times N$ matrix
is truncated to the first three rows.
The Majorana phases
$\lambda_{21}, \ldots \lambda_{N1}$,
which are physical only if massive neutrinos are Majorana particles,
are collected in a diagonal matrix on the right\footnote{
It is possible to choose any other diagonal matrix with $N-1$ phases,
as for example
$
\operatorname{diag}\!\left(
e^{i\lambda_{12}}, 1, e^{i\lambda_{32}}, \ldots, e^{i\lambda_{N2}}
\right)
$,
etc.
}.
Moreover,
not all the phases $\eta_{ab}$ in the product of $W^{ab}$ matrices in Eq.~(\ref{f470})
are physical, but one can eliminate an unphysical phase for each value of the
index $b=4,\ldots,N$
(see Ref.~\cite{Giunti:2007ry}).

The scheme (\ref{f470})
has the advantage that in the limit of vanishing active-sterile mixing the
mixing matrix reduces to the three-neutrino ($3\nu$) mixing matrix in the standard parameterization
\begin{align}
\null & \null
U^{(3\nu)}
=
\left[
R^{23}
W^{13}
R^{12}
\right]_{3 \times 3}
\operatorname{diag}\!\left(
1, e^{i\lambda_{21}}, e^{i\lambda_{31}}
\right)
\nonumber
\\
=
\null & \null
\begin{pmatrix}
c_{12}
c_{13}
&
s_{12}
c_{13}
&
s_{13}
e^{-i\eta_{13}}
\\
-
s_{12}
c_{23}
-
c_{12}
s_{23}
s_{13}
e^{i\eta_{13}}
&
c_{12}
c_{23}
-
s_{12}
s_{23}
s_{13}
e^{i\eta_{13}}
&
s_{23}
c_{13}
\\
s_{12}
s_{23}
-
c_{12}
c_{23}
s_{13}
e^{i\eta_{13}}
&
-
c_{12}
s_{23}
-
s_{12}
c_{23}
s_{13}
e^{i\eta_{13}}
&
c_{23}
c_{13}
\end{pmatrix}
\begin{pmatrix}
1 & 0 & 0
\\
0 & e^{i\lambda_{21}} & 0
\\
0 & 0 & e^{i\lambda_{31}}
\end{pmatrix}
.
\label{3numix}
\end{align}

It is convenient to choose
in Eq.~(\ref{f470})
the order of the real or complex rotations for each index $b \geq 4$
such that the rotations in the
3-$b$,
2-$b$ and
1-$b$ planes
are ordered from left to right.
In this way,
the first two lines,
which are relevant for the study of the oscillations of
the experimentally more accessible flavor neutrinos $\nu_{e}$ and $\nu_{\mu}$,
are independent of the mixing angles and Dirac phases
corresponding to the rotations
in all the 3-$b$ planes for $b \geq 4$.
Moreover,
the first line,
which is relevant for the study of $\nu_{e}$ disappearance,
is independent also of the mixing angles and Dirac phases
corresponding to the rotations
in the 2-$b$ planes for $b \geq 3$.
For example,
one can choose
\begin{equation}
U
=
\left[
W^{3N}
R^{2N}
W^{1N}
\cdots
W^{34}
R^{24}
W^{14}
R^{23}
W^{13}
R^{12}
\right]_{3 \times N}
\operatorname{diag}\!\left(
1, e^{i\lambda_{21}}, \ldots, e^{i\lambda_{N1}}
\right)
,
\label{examix1}
\end{equation}
or
\begin{equation}
U
=
\left[
W^{3N}
\cdots
W^{34}
W^{2N}
\cdots
W^{24}
R^{1N}
\cdots
R^{14}
R^{23}
W^{13}
R^{12}
\right]_{3 \times N}
\operatorname{diag}\!\left(
1, e^{i\lambda_{21}}, \ldots, e^{i\lambda_{N1}}
\right)
.
\label{examix2}
\end{equation}

Let us now consider the neutrino neutral-current Lagrangian
\begin{equation}
\mathcal{L}_{\text{NC}}
=
-
\frac{ g }{ 2 \cos\vartheta_{\text{W}} }
\overline{\nu^{(\text{a})}_{L}} \gamma^{\rho} \nu^{(\text{a})}_{L} Z_{\rho}
=
-
\frac{ g }{ 2 \cos\vartheta_{\text{W}} }
\overline{\nu^{(\text{M})}_{L}} \gamma^{\rho} U^{\dagger} U \nu^{(\text{M})}_{L} Z_{\rho}
.
\label{208}
\end{equation}
Since the rectangular
$3 \times N$
mixing matrix $U$ is formed by the first three rows of the unitary matrix $\mathcal{U}$,
we have
\begin{equation}
U U^{\dagger} = \openone_{3\times3}
,
\qquad
\text{but}
\qquad
U^{\dagger} U \neq \openone_{N \times N}
.
\label{209}
\end{equation}
Therefore, the GIM mechanism \cite{Glashow:1970gm}
does not work in neutral-current weak interactions \cite{Schechter:1980gr}
and it is possible to have neutral-current transitions among different massive neutrinos\footnote{
This is a special case of
the general theorem that the weak leptonic neutral current is nondiagonal
in the mass basis if the leptons of a given charge and chirality have
different weak isospins \cite{Lee:1977tib}.
}.

The introduction of sterile neutrinos is allowed by the fact that
it has no effect or small effects\footnote{
Sterile neutrinos at mass scales larger than the muon mass affect the determination of the Fermi constant
$G_{\text{F}}$
through muon decay
\cite{Antusch:2006vwa}.
Those at mass scales larger than $m_{Z}/2$
can induce a kinematical suppression of $N_{\nu}^{(Z)}$
\cite{Jarlskog:1990kt,Bilenky:1990tm}.
}
on the effective number of active neutrinos
which contributes to the decay of the $Z$-boson.
This number has been determined with high precision to be close to three by the LEP experiments
\cite{ALEPH:2005ab}:
\begin{equation}
N_{\nu}^{(Z)}
=
2.9840 \pm 0.0082
.
\label{211}
\end{equation}
In this review we will consider sterile neutrinos at the eV scale,
for which
$N_{\nu}^{(Z)}$
is given by
\cite{Jarlskog:1990kt,Bilenky:1990tm}
\begin{equation}
N_{\nu}^{(Z)}
=
\sum_{j,k=1}^{N}
\left|
\sum_{\alpha=e,\mu,\tau}
U_{\alpha j}^{*}
\,
U_{\alpha k}
\right|^2
=
3
.
\label{212}
\end{equation}
Hence,
there is no constraint
on the number and mixing of these light sterile neutrinos
from the high-precision LEP measurement of $N_{\nu}^{(Z)}$.

The measurements of neutrino oscillations in solar, reactor and accelerator experiments
determined the existence of two squared-mass differences:
the solar and atmospheric squared-mass differences
\begin{equation}
\Delta m^2_{\text{SOL}}
\simeq
7.5 \times 10^{-5} \, \text{eV}^2
,
\qquad
\Delta m^2_{\text{ATM}}
\simeq
2.4 \times 10^{-3} \, \text{eV}^2
.
\label{213}
\end{equation}
It is convenient to label the three light neutrino masses
according to the convention
\begin{equation}
\Delta{m}^{2}_{\text{SOL}}
=
\Delta{m}^{2}_{21}
\ll
\Delta{m}^{2}_{\text{ATM}}
=
\frac{1}{2}
\left|
\Delta{m}^{2}_{31}
+
\Delta{m}^{2}_{32}
\right|
,
\label{B081}
\end{equation}
with
$\Delta{m}^{2}_{jk} = m_{j}^2 - m_{k}^2$.
The absolute value in the definition of $\Delta{m}^{2}_{\text{ATM}}$
is necessary,
because
there are two possible orderings of the neutrino masses:
the normal ordering
(NO)
with
$m_{1}<m_{2}<m_{3}$
and
$\Delta{m}^{2}_{31}, \, \Delta{m}^{2}_{32} > 0$;
the inverted ordering
(IO)
with
$m_{3}<m_{1}<m_{2}$
and
$\Delta{m}^{2}_{31}, \, \Delta{m}^{2}_{32} < 0$.

Table~\ref{tab:global3nu}
shows the results of the determination of the $3\nu$ mixing parameters
obtained in Ref.~\cite{Capozzi:2013csa}
from a global fit of neutrino oscillation data
(see also Refs.~\cite{Forero:2014bxa,Gonzalez-Garcia:2014bfa}).
The largest uncertainty is that of $\vartheta_{23}$,
which is known to be close to maximal ($\pi/4$),
but it is not known if it is smaller or larger than $\pi/4$.
For the Dirac CP-violating phase $\eta_{13}$,
there is an indication in favor of $\eta_{13} \approx 3\pi/2$,
which would give maximal CP violation,
but at $3\sigma$ all the values of $\eta_{13}$ are allowed,
including the CP-conserving values $\eta_{13}=0,\pi$.

\begin{table}[t]
\begin{center}
\begin{tabular}{lccccc}
parameter
&
\begin{tabular}{c}
mass
\\[-0.1cm]
order
\end{tabular}
&
\begin{tabular}{c}
best
\\[-0.1cm]
fit
\end{tabular}
&
$1\sigma$ range
&
$2\sigma$ range
&
$3\sigma$ range
\\
\hline
$\Delta{m}^2_{\text{SOL}}/10^{-5}\,\text{eV}^2 $ & & 7.54 & 7.32 -- 7.80 & 7.15 -- 8.00 & 6.99 -- 8.18 \\
\hline
$\sin^2 \vartheta_{12}/10^{-1}$ & & 3.08 & 2.91 -- 3.25 & 2.75 -- 3.42 & 2.59 -- 3.59 \\
\hline
\multirow{2}{*}{$\Delta{m}^2_{\text{ATM}}/10^{-3}\,\text{eV}^2$}
& NO & 2.43 & 2.37 -- 2.49 & 2.30 -- 2.55 & 2.23 -- 2.61 \\
& IO & 2.38 & 2.32 -- 2.44 & 2.25 -- 2.50 & 2.19 -- 2.56 \\
\hline
\multirow{2}{*}{$\sin^2 \vartheta_{23}/10^{-1}$}
& NO & 4.37 & 4.14 -- 4.70 & 3.93 -- 5.52 & 3.74 -- 6.26 \\
& IO & 4.55 & 4.24 -- 5.94 & 4.00 -- 6.20 & 3.80 -- 6.41 \\
\hline
\multirow{2}{*}{$\sin^2 \vartheta_{13}/10^{-2}$}
& NO & 2.34 & 2.15 -- 2.54 & 1.95 -- 2.74 & 1.76 -- 2.95 \\
& IO & 2.40 & 2.18 -- 2.59 & 1.98 -- 2.79 & 1.78 -- 2.98 \\
\hline
\end{tabular}
\end{center}
\caption{\label{tab:global3nu}
Values of the neutrino mixing parameters obtained in Ref.~\cite{Capozzi:2013csa} with a
global analysis of neutrino oscillation data
in the framework of three-neutrino mixing
with the normal ordering (NO) and the inverted ordering (IO).
}
\end{table}

The standard framework of $3\nu$ mixing
can be extended with the introduction of non-standard massive neutrinos
only if their mixing with the active neutrinos is sufficiently small
in order not to spoil the successful $3\nu$ mixing explanation of
solar, atmospheric and long-baseline
neutrino oscillation measurements discussed above.
In other words,
the non-standard massive neutrinos must be mostly sterile,
i.e.
\begin{equation}
|U_{\alpha k}|^2 \ll 1
\qquad
(\alpha=e,\mu,\tau; \, k=4, \ldots, N)
.
\label{smallmix}
\end{equation}
In the following we will always assume this constraint.

\begin{figure}[t]
\begin{center}
\setlength{\tabcolsep}{0.5cm}
\begin{tabular}{cccccc}
\includegraphics*[width=0.09\textwidth]{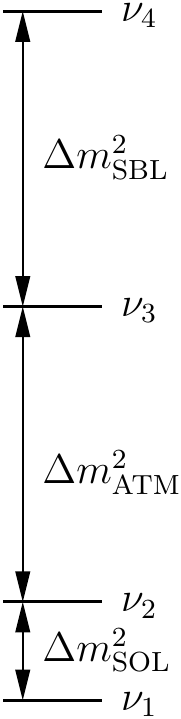}
&
\includegraphics*[width=0.09\textwidth]{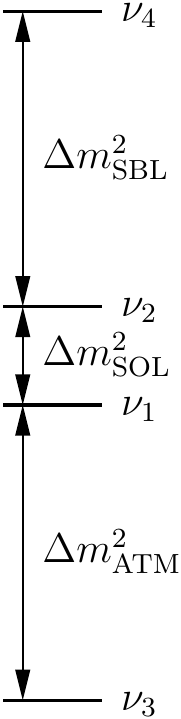}
&
\includegraphics*[width=0.09\textwidth]{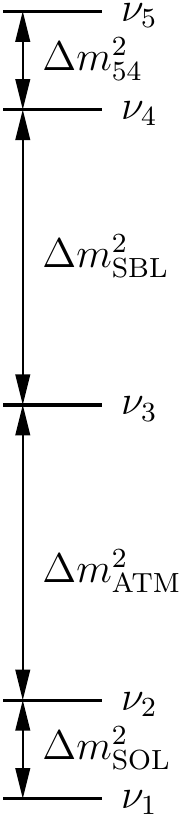}
&
\includegraphics*[width=0.09\textwidth]{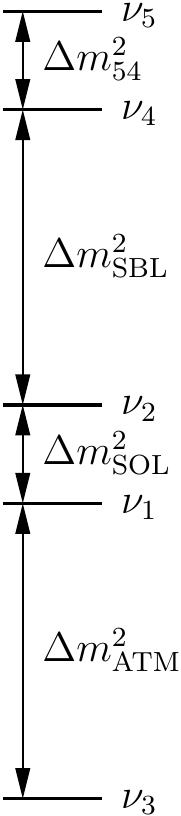}
&
\includegraphics*[width=0.09\textwidth]{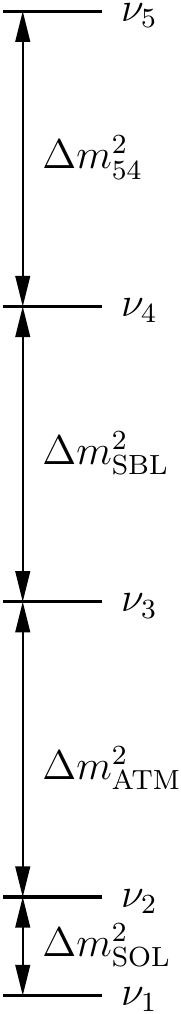}
&
\includegraphics*[width=0.09\textwidth]{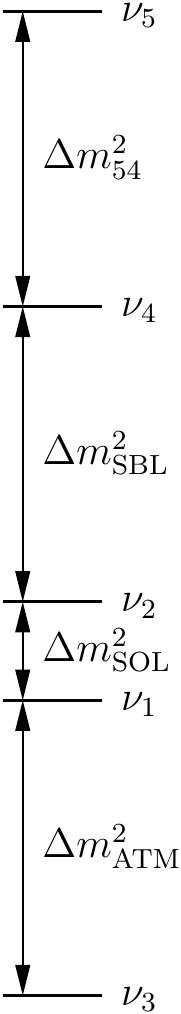}
\\
$3_{\text{NO}}$+1
&
$3_{\text{IO}}$+1
&
$3_{\text{NO}}$+2
&
$3_{\text{IO}}$+2
&
$3_{\text{NO}}$+1+1
&
$3_{\text{IO}}$+1+1
\end{tabular}
\end{center}
\caption{\label{fig:schemes}
Schematic illustration of the
3+1, 3+2 and 3+1+1 neutrino mixing schemes
taking into account for each scheme the two
possible mass ordering,
normal (NO) and inverted (IO),
of the three lightest standard neutrinos.
In the two 3+2 schemes
$\Delta m^2_{54} \approx \Delta m^2_{\text{SBL}}$,
whereas in the two 3+1+1 schemes
$\Delta m^2_{54} \gg \Delta m^2_{\text{SBL}}$.
}
\end{figure}

In this review we consider mainly the so-called 3+1 scheme
in which there is a non-standard massive neutrino (mostly sterile) at the eV scale
which generates a new squared-mass difference
\begin{equation}
\Delta m^2_{\text{SBL}}
\sim
1 \, \text{eV}^2
,
\label{dm2sbl}
\end{equation}
in order to
explain the anomalies found in some short-baseline (SBL) neutrino oscillation experiments
(see Section~\ref{sec:sbl}).
We assume that the three standard massive neutrinos are much lighter than the eV scale.
We will consider also the so-called 3+2 scheme
in which there are two non-standard massive neutrinos (mostly sterile) at the eV scale
and the 3+1+1 scheme in which there is a non-standard massive neutrino at the eV scale
and another at a larger scale.
We do not consider schemes in which $\Delta m^2_{\text{SBL}}$
is obtained with one or more very light (or massless)
non-standard massive neutrinos and the three standard massive neutrinos have almost degenerate masses at the eV scale
(e.g., the 1+3, 1+3+1 and 2+3 schemes),
because this possibility is strongly disfavored by cosmological measurements
\cite{Ade:2015xua}
and by the experimental bound on
neutrinoless double-$\beta$ decay
(assuming that massive neutrinos are Majorana particles;
see Ref.~\cite{Bilenky:2014uka}).
Figure~\ref{fig:schemes}
shows a schematic illustration of the
3+1, 3+2 and 3+1+1 neutrino mixing schemes
taking into account for each scheme the two
possible mass orderings
of the three lightest standard neutrinos.
Let us emphasize that these mixing schemes must be considered as effective,
in the sense that the existence of more non-standard massive neutrinos
is allowed, as long as their mixing with the three active neutrinos is sufficiently small
to be negligible in the analysis of the data of current experiments.

For the study of neutrino oscillations in vacuum it is convenient to use
the following general expression of the probability of
$\nua{\alpha}\to\nua{\beta}$
oscillations
\cite{Bilenky:2012zp,Bilenky:2015xwa}:
%
%
\begin{align}
P_{\nua{\alpha}\to\nua{\beta}}
=
\null & \null
\delta_{\alpha\beta}
-
4
\sum_{k \neq p}
|U_{\alpha k}|^2
\left( \delta_{\alpha\beta} -  |U_{\beta k}|^2 \right)
\sin^{2}\Delta_{kp}
\nonumber
\\
\null & \null
+
8
\sum_{\stackrel{\scriptstyle j>k}{\scriptstyle j,k \neq p}}
\left|
U_{\alpha j}
U_{\beta j}
U_{\alpha k}
U_{\beta k}
\right|
\sin\Delta_{kp}
\sin\Delta_{jp}
\cos(
\Delta_{jk}
\stackrel{(+)}{-}
\eta_{\alpha\beta jk}
)
,
\label{221}
\end{align}
where
\begin{equation}
\Delta_{kp} = \frac{ \Delta{m}^{2}_{kp} L }{ 4 E }
,
\qquad
\eta_{\alpha\beta jk}
=
\operatorname{arg}\!\left[
U_{\alpha j}^{*}
U_{\beta j}
U_{\alpha k}
U_{\beta k}^{*}
\right]
,
\label{222}
\end{equation}
and $p$ is an arbitrary fixed index,
which can be chosen in the most convenient way
depending on the case under consideration.
In the case of three-neutrino mixing,
there is only one interference term in Eq.~(\ref{221}),
because for any choice of $p$ there is only one possibility for $j$ and $k$
such that $j>k$.

We are interested in the effective oscillation probabilities
in short-baseline experiments,
for which
$\Delta_{21} \ll \Delta_{31} \ll 1$.
Let us consider the general
3+$N_{s}$
case in which
$\Delta{m}^2_{k1} \approx \Delta{m}^2_{\text{SBL}}$
and
$\Delta_{k1} \approx 1$
for $k\geq4$.
Choosing $p=1$ in Eq.~(\ref{222}),
we obtain
\begin{align}
P_{\nua{\alpha}\to\nua{\beta}}^{(\text{SBL})}
\simeq
\null & \null
\delta_{\alpha\beta}
-
4
\sum_{k=4}^{N}
|U_{\alpha k}|^2
\left( \delta_{\alpha\beta} -  |U_{\beta k}|^2 \right)
\sin^{2}\Delta_{k1}
\nonumber
\\
\null & \null
+
8
\sum_{k=4}^{N}
\sum_{j=k+1}^{N}
\left|
U_{\alpha j}
U_{\beta j}
U_{\alpha k}
U_{\beta k}
\right|
\sin\Delta_{k1}
\sin\Delta_{j1}
\cos(
\Delta_{jk}
\stackrel{(+)}{-}
\eta_{\alpha\beta jk}
)
.
\label{223}
\end{align}

Considering the survival probabilities of active neutrinos,
let us define the effective amplitudes
\begin{equation}
\sin^2 2\vartheta_{\alpha\alpha}^{(k)}
=
4 |U_{\alpha k}|^2 \left( 1 - |U_{\alpha k}|^2 \right)
\simeq
4 |U_{\alpha k}|^2
\qquad
(\alpha=e,\mu,\tau; \, k\geq4)
,
\label{asurv}
\end{equation}
where we have taken into account the constraint in Eq.~(\ref{smallmix}).
Dropping the quadratically suppressed terms also in the survival probabilities,
we obtain
\begin{equation}
P_{\nua{\alpha}\to\nua{\alpha}}^{(\text{SBL})}
\simeq
1
-
\sum_{k=4}^{N}
\sin^2 2\vartheta_{\alpha\alpha}^{(k)}
\sin^{2}\Delta_{k1}
\qquad
(\alpha=e,\mu,\tau)
.
\label{psurv}
\end{equation}
Hence,
each effective mixing angle
$\vartheta_{\alpha\alpha}^{(k)}$
parameterizes the disappearance of $\nua{\alpha}$
due to its mixing with $\nua{k}$.

Let us now consider the probabilities of short-baseline
$\nua{\alpha}\to\nua{\beta}$
transitions between two different active neutrinos
or an active and a sterile neutrino.
We define the transition amplitudes
\begin{equation}
\sin^2 2\vartheta_{\alpha\beta}^{(k)}
=
4 |U_{\alpha k}|^2 |U_{\beta k}|^2
\qquad
(\alpha\neq\beta; \, k\geq4)
,
\label{atran}
\end{equation}
which allow us to write the transition probabilities as
\begin{align}
P_{\nua{\alpha}\to\nua{\beta}}^{(\text{SBL})}
\simeq
\null & \null
\sum_{k=4}^{N}
\sin^2 2\vartheta_{\alpha\beta}^{(k)}
\sin^{2}\Delta_{k1}
\nonumber
\\
\null & \null
+
2
\sum_{k=4}^{N}
\sum_{j=k+1}^{N}
\sin 2\vartheta_{\alpha\beta}^{(k)}
\sin 2\vartheta_{\alpha\beta}^{(j)}
\sin\Delta_{k1}
\sin\Delta_{j1}
\cos(
\Delta_{jk}
\stackrel{(+)}{-}
\eta_{\alpha\beta jk}
)
.
\label{ptran}
\end{align}
From the first line
one can see that
each effective mixing angle
$\vartheta_{\alpha\beta}^{(k)}$
parameterizes the amount of
$\nua{\alpha}\to\nua{\beta}$
transitions
due to the mixing of $\nua{\alpha}$ and $\nua{\beta}$ with $\nua{k}$.
The second line in Eq.~(\ref{ptran})
is the interference between the $\nua{k}$ and $\nua{j}$
contributions,
which depends on the same effective mixing angles.

Considering now the transitions between two different active neutrinos,
from Eqs.~(\ref{asurv}) and (\ref{atran})
one can see that for each value of $k\geq4$
the transition amplitude
$\sin 2\vartheta_{\alpha\beta}^{(k)}$
and
the disappearance amplitudes
$\sin 2\vartheta_{\alpha\alpha}^{(k)}$
and
$\sin 2\vartheta_{\beta\beta}^{(k)}$
depend only on the elements in $k^{\text{th}}$ column of the mixing matrix
and are related by\footnote{
This relation was derived in the case of 3+1 mixing
(see Eq.~(\ref{appdis3p1}))
in Refs.~\cite{Okada:1996kw,Bilenky:1996rw}.
}
\cite{Giunti:2015mwa}
\begin{equation}
\sin^2 2\vartheta_{\alpha\beta}^{(k)}
\simeq
\frac{1}{4}
\,
\sin^2 2\vartheta_{\alpha\alpha}^{(k)}
\,
\sin^2 2\vartheta_{\beta\beta}^{(k)}
\qquad
(\alpha=e,\mu,\tau)
.
\label{appdis}
\end{equation}
This relation is very important,
because it constrains the oscillation signals that can be observed in
short-baseline appearance and disappearance experiments
in any 3+$N_{s}$ mixing scheme with sterile neutrinos.
Its experimental test is crucial for the acceptance or rejection of
these schemes.
In particular,
since both
$\sin^2 2\vartheta_{\alpha\alpha}^{(k)}$
and
$\sin^2 2\vartheta_{\beta\beta}^{(k)}$
are small for $\alpha,\beta=e,\mu,\tau$
the amplitudes of the short-baseline transition probabilities
between active neutrinos are quadratically suppressed.
We will see in Section~\ref{sec:global}
that the current short-baseline data
have an appearance-disappearance
tension due to the constraint in Eq.~(\ref{appdis}).

In the following part of this Section we discuss briefly the main
peculiar characteristics of short-baseline oscillations
in the cases of
3+1,
3+2 and
3+1+1 neutrino mixing.

\noindent
\textbf{3+1 mixing.}
In the case of 3+1 neutrino mixing
\cite{Okada:1996kw,Bilenky:1996rw,Bilenky:1999ny,Maltoni:2004ei},
we have
$\Delta{m}^2_{41} = \Delta{m}^2_{\text{SBL}}$
and
$\Delta_{41} \sim 1$
in short-baseline experiments.
The transition and survival probabilities
can be written as
\begin{equation}
P_{\nua{\alpha}\to\nua{\beta}}^{(\text{SBL})}
\simeq
\sin^2 2\vartheta_{\alpha\beta}
\sin^{2}\Delta_{41}
\quad
(\alpha\neq\beta)
,
\qquad
P_{\nua{\alpha}\to\nua{\alpha}}^{(\text{SBL})}
\simeq
1
-
\sin^2 2\vartheta_{\alpha\alpha}
\sin^{2}\Delta_{41}
,
\label{pro3p1}
\end{equation}
with the transition and survival amplitudes
\begin{equation}
\sin^2 2\vartheta_{\alpha\beta}
=
4 |U_{\alpha 4}|^2 |U_{\beta 4}|^2
\quad
(\alpha\neq\beta)
,
\qquad
\sin^2 2\vartheta_{\alpha\alpha}
=
4
|U_{\alpha 4}|^2
\left(1 -  |U_{\alpha 4}|^2 \right)
,
\label{amp3p1}
\end{equation}
and with the appearance-disappearance constraint
\cite{Okada:1996kw,Bilenky:1996rw}
\begin{equation}
\sin^2 2\vartheta_{\alpha\beta}
\simeq
\frac{1}{4}
\,
\sin^2 2\vartheta_{\alpha\alpha}
\,
\sin^2 2\vartheta_{\beta\beta}
\qquad
(\alpha=e,\mu,\tau)
.
\label{appdis3p1}
\end{equation}

The transition and survival
probabilities in Eq.~(\ref{pro3p1})
depend only on the largest squared-mass difference
$\Delta{m}^2_{41} = \Delta{m}^2_{\text{SBL}}$
and on the absolute values of the elements in the fourth column of the mixing matrix.
The transition probabilities of neutrinos and antineutrinos are equal,
because the absolute values of the elements in the fourth column of the mixing matrix
do not depend on the CP-violating phases in the mixing matrix.
Hence, even if there are CP-violating phases in the mixing matrix,
CP violation cannot be measured in short-baseline experiments.
In order to measure the effects of these phases
it is necessary to perform experiments sensitive to the oscillations generated by the smaller squared-mass differences
$\Delta{m}^2_{\text{ATM}}$
\cite{deGouvea:2014aoa,Klop:2014ima,Berryman:2015nua}
or
$\Delta{m}^2_{\text{SOL}}$
\cite{Long:2013hwa}.

\noindent
\textbf{3+2 mixing.}
In the case of 3+2 neutrino mixing
\cite{Sorel:2003hf,Karagiorgi:2006jf,Maltoni:2007zf,Karagiorgi:2009nb,Blennow:2011vn},
we have
$\Delta{m}^2_{51} \approx \Delta{m}^2_{41} = \Delta{m}^2_{\text{SBL}}$
and
$\Delta_{51} \approx \Delta_{41} \sim 1$
in short-baseline experiments.
From Eq.~(\ref{psurv})
we obtain the short-baseline survival probabilities of active neutrinos
\begin{equation}
P_{\nua{\alpha}\to\nua{\alpha}}^{(\text{SBL})}
\simeq
1
-
\sin^2 2\vartheta_{\alpha\alpha}^{(4)}
\sin^{2}\Delta_{41}
-
\sin^2 2\vartheta_{\alpha\alpha}^{(5)}
\sin^{2}\Delta_{51}
\qquad
(\alpha=e,\mu,\tau)
,
\label{psurv3p2}
\end{equation}
and from Eq.~(\ref{ptran})
we obtain the short-baseline transition probabilities
\begin{align}
P_{\nua{\alpha}\to\nua{\beta}}^{(\text{SBL})}
\simeq
\null & \null
\sin^2 2\vartheta_{\alpha\beta}^{(4)}
\sin^{2}\Delta_{41}
+
\sin^2 2\vartheta_{\alpha\beta}^{(5)}
\sin^{2}\Delta_{51}
\nonumber
\\
\null & \null
+
2
\sin 2\vartheta_{\alpha\beta}^{(4)}
\sin 2\vartheta_{\alpha\beta}^{(5)}
\sin\Delta_{41}
\sin\Delta_{51}
\cos(
\Delta_{54}
\stackrel{(+)}{-}
\eta_{\alpha\beta 54}
)
\qquad
(\alpha\neq\beta)
.
\label{ptran3p2}
\end{align}
The appearance and disappearance amplitudes are related by the general constraint in Eq.~(\ref{appdis}).
The 3+2 scheme has the important characteristic that
CP violation is observable in short-baseline experiments
through the asymmetries
\begin{equation}
A_{\alpha\beta}^{(\text{SBL})}
=
P_{\nu_{\alpha}\to\nu_{\beta}}^{(\text{SBL})}
-
P_{\bar\nu_{\alpha}\to\bar\nu_{\beta}}^{(\text{SBL})}
\simeq
4
\sin 2\vartheta_{\alpha\beta}^{(4)}
\sin 2\vartheta_{\alpha\beta}^{(5)}
\sin\Delta_{41}
\sin\Delta_{51}
\sin\Delta_{54}
\sin\eta_{\alpha\beta 54}
,
\label{acp3p2}
\end{equation}
for $\alpha\neq\beta$.

\noindent
\textbf{3+1+1 mixing.}
In the case of 3+1+1 mixing
\cite{Nelson:2010hz,Fan:2012ca,Kuflik:2012sw,Huang:2013zga},
we have
$\Delta{m}^2_{51} \gg \Delta{m}^2_{41} = \Delta{m}^2_{\text{SBL}}$
and
$\Delta_{51} \gg \Delta_{41} \sim 1$
in short-baseline experiments.
The corresponding oscillation probabilities can be obtained from those in the case of
3+2 mixing
by averaging the oscillations due to
$\Delta{m}^2_{51}$:
\begin{equation}
P_{\nua{\alpha}\to\nua{\alpha}}^{(\text{SBL})}
\simeq
1
-
\sin^2 2\vartheta_{\alpha\alpha}^{(4)}
\sin^{2}\Delta_{41}
-
\frac{1}{2}
\,
\sin^2 2\vartheta_{\alpha\alpha}^{(5)}
\qquad
(\alpha=e,\mu,\tau)
,
\label{psurv3p1p1}
\end{equation}
and
\begin{align}
P_{\nua{\alpha}\to\nua{\beta}}^{(\text{SBL})}
\simeq
\null & \null
\sin^2 2\vartheta_{\alpha\beta}^{(4)}
\sin^{2}\Delta_{41}
+
\frac{1}{2}
\,
\sin^2 2\vartheta_{\alpha\beta}^{(5)}
\nonumber
\\
\null & \null
+
\sin 2\vartheta_{\alpha\beta}^{(4)}
\sin 2\vartheta_{\alpha\beta}^{(5)}
\sin\Delta_{41}
\sin(
\Delta_{41}
\stackrel{(-)}{+}
\eta_{\alpha\beta 54}
)
\qquad
(\alpha\neq\beta)
.
\label{ptran3p1p1}
\end{align}
Hence,
in the analysis of short-baseline data
in the 3+1+1 scheme
there is one effective parameter less than in the 3+2 scheme
($\Delta{m}^2_{51}$),
but CP violation effects generated by the phases $\eta_{\alpha\beta 54}$ are observable.
\section{Short-baseline anomalies and constraints}
\label{sec:sbl}

In this Section we review the three experimental indications
in favor of short-baseline neutrino oscillations,
which require the existence of at least one additional squared-mass difference,
$\Delta{m}^2_{\text{SBL}}$,
which is much larger than
$\Delta{m}^2_{\text{SOL}}$
and
$\Delta{m}^2_{\text{ATM}}$:
the reactor antineutrino anomaly in Subsection~\ref{sub:reactor},
the Gallium neutrino anomaly in Subsection~\ref{sub:gallium},
and
the LSND anomaly in Subsection~\ref{sub:LSND}.

\subsection{The reactor antineutrino anomaly}
\label{sub:reactor}


\begin{figure}[t]
\begin{center}
\includegraphics*[width=\textwidth]{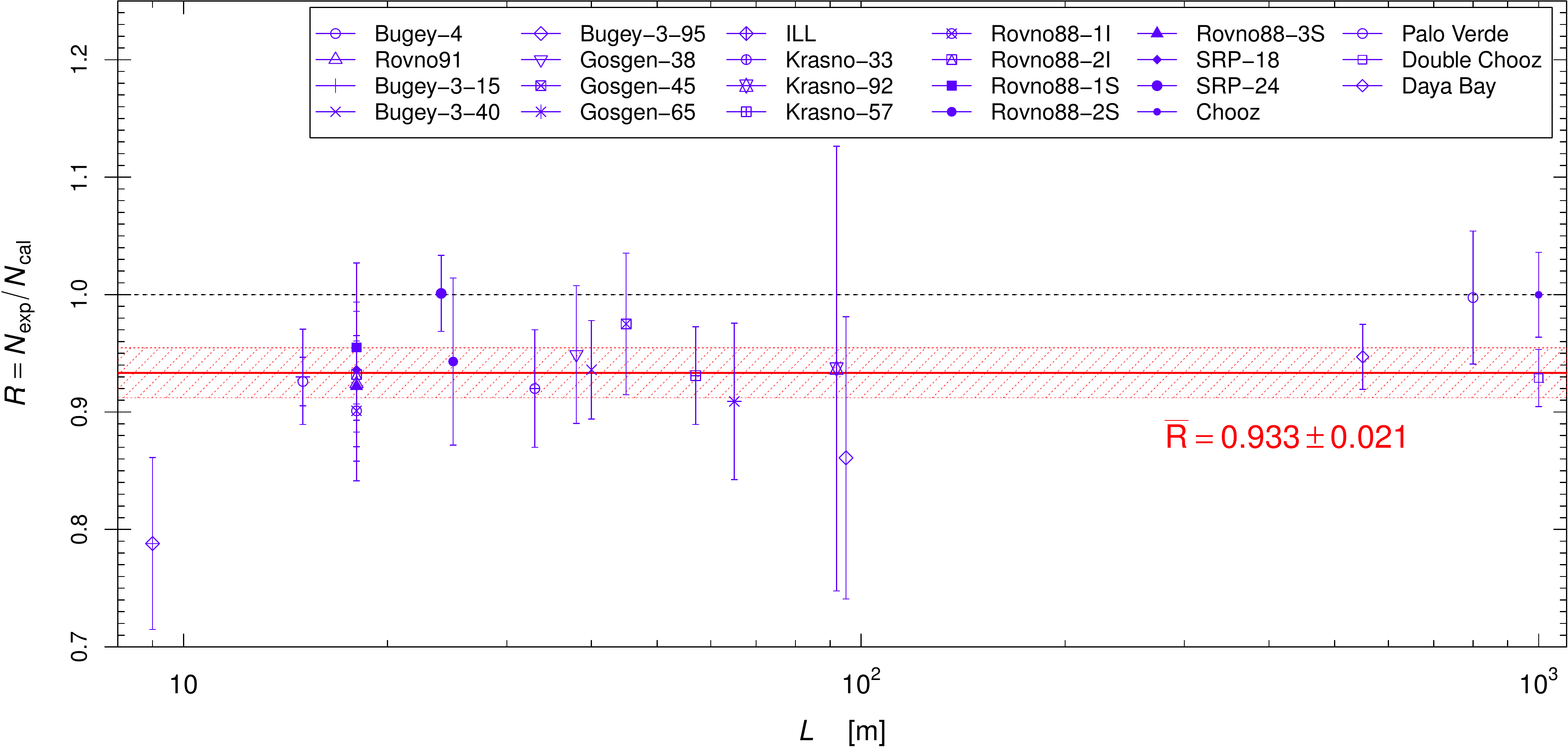}
\end{center}
\caption{ \label{fig:rea-lst}
Ratios $R$ of the measured
($N_{\text{exp}}$)
and calculated
($N_{\text{cal}}$)
number of electron antineutrino events in reactor experiments at different distances $L$.
The horizontal shadowed red band shows the average ratio $\overline{R}$ and its uncertainty.
For each experiment
the error bar shows the experimental uncertainty.
The values of the ratios of the long-baseline experiments
Daya Bay, Double Chooz, Chooz and Palo Verde
have been obtained by subtracting the effect of $\vartheta_{13}$-driven oscillations.
}
\end{figure}

The reactor antineutrino anomaly
\cite{Mention:2011rk},
is a deficit of the rate of $\bar\nu_{e}$ observed in several
short-baseline reactor neutrino experiments
in comparison with that expected from the calculation of
the reactor neutrino fluxes
\cite{Mueller:2011nm,Huber:2011wv,Abazajian:2012ys}.
It has been analyzed by several authors
\cite{Mention:2011rk,Sinev:2011ra,Kopp:2011qd,Giunti:2011gz,Giunti:2011hn,Giunti:2011cp,Abazajian:2012ys,Ciuffoli:2012yd,Giunti:2012tn,Giunti:2012bc,Zhang:2013ela,Kopp:2013vaa,Ivanov:2013cga,Giunti:2013aea,Girardi:2014wea}.

The statistical significance of the anomaly depends on the estimated uncertainties
of the calculated reactor antineutrino fluxes,
on which there have been some debate
\cite{Fallot:2012jv,Hayes:2013wra,Dwyer:2014eka,Ma:2014bpa,Sonzogni:2015aoa,Fang:2015cma,Zakari-Issoufou:2015vvp,Hayes:2015yka},
especially after the discovery of an excess at about 5 MeV of the reactor antineutrino spectrum
in the
RENO \cite{Seon-HeeSeofortheRENO:2014jza},
Double Chooz \cite{Crespo-Anadon:2014dea}
and
Daya Bay \cite{Zhang:2015fya,Zhan:2015aha,An:2015nua}
experiments.
However,
since this problem is controversial and far from being solved,
in the following we assume the
values and uncertainties of the reactor antineutrino fluxes
presented in Refs.~\cite{Mueller:2011nm,Huber:2011wv,Abazajian:2012ys},
which have been obtained from the available database information on nuclear decays
and from the electron spectra associated with the fission of
$^{235}\text{U}$,
$^{239}\text{Pu}$, and
$^{241}\text{Pu}$
measured at ILL in the 80's
\cite{VonFeilitzsch:1982jw,Schreckenbach:1985ep,Hahn:1989zr,Haag:2014kia}.
The electron spectra associated with the fission of
$^{235}\text{U}$
and
$^{238}\text{U}$
has been measured recently at the scientific neutron source FRM II in Garching
\cite{Haag:2013raa}.
In this experiment the $^{235}\text{U}$ electron spectrum has been normalized to that measured at ILL \cite{Schreckenbach:1985ep,Haag:2014kia}
in order to reduce the systematic uncertainties.
The $^{238}\text{U}$ antineutrino spectrum obtained from the conversion
has only a spectral distortion of about 10\% with respect to the
standard $^{238}\text{U}$ antineutrino spectrum
\cite{Mueller:2011nm,Huber:2011wv,Abazajian:2012ys}.

In reactor neutrino experiments
electron antineutrinos are
detected through the inverse neutron decay process
\begin{equation}
\bar\nu_{e} + p \to n + e^{+}
\label{l011}
\end{equation}
in liquid-scintillator detectors.
The cross section of the detection process is
$
\sigma_{\bar\nu_{e}p}(E_{e})
\propto
E_{e} p_{e}
$
(see Refs.~\cite{Bemporad:2001qy,Fukugita:2003en,Giunti:2007ry}),
where $E_{e}$ and $p_{e}$ are, respectively, the positron energy and momentum.
Neglecting the small recoil energy of the neutron,
the neutrino energy $E$ can be calculated from
the measurable kinetic energy $T_{e}$ of the positron through the relation
\begin{equation}
E
\simeq
T_{e} + m_{e} + m_{n} - m_{p}
\simeq
T_{e} + 1.8 \, \text{MeV}
,
\label{l013}
\end{equation}
where $m_{p}$ and $m_{n}$ are, respectively, the proton and neutron masses.
Hence, the neutrino energy threshold for the detection process is about $1.8 \, \text{MeV}$.

Figure~\ref{fig:rea-lst} shows the
ratios $R$ of the measured
($N_{\text{exp}}$)
and calculated
($N_{\text{cal}}$)
number of electron antineutrino events in the reactor experiments
Bugey-4 \cite{Declais:1994ma},
ROVNO91 \cite{Kuvshinnikov:1990ry},
Bugey-3 \cite{Declais:1995su},
Gosgen \cite{Zacek:1986cu},
ILL \cite{Hoummada:1995zz},
Krasnoyarsk \cite{Vidyakin:1990iz},
Rovno88 \cite{Afonin:1988gx},
SRP \cite{Greenwood:1996pb},
Chooz \cite{Apollonio:2002gd},
Palo Verde \cite{Boehm:2001ik},
Double Chooz \cite{Abe:2014bwa}, and
Daya Bay \cite{Zhang:2015fya,Zhan:2015aha,An:2015nua}
at different distances $L$.
As shown in the figure,
the average ratio is
$\overline{R} = 0.933 \pm 0.021$,
which indicates a deficit with a nominal\footnote{
We call ``nominal'' the statistical significances of the indications
in favor of short-baseline neutrino oscillations
in order to emphasize that they depend on the estimated standard uncertainties.
}
statistical significance of about
$3.1\sigma$.

\begin{table}[t]
\begin{center}
\begin{tabular}{l|cccc|cc}
&
\multicolumn{4}{c|}{${}^{51}\text{Cr}$}
&
\multicolumn{2}{c}{${}^{37}\text{Ar}$}
\\
\hline
$E\,[\text{keV}]$		& $  747 $ & $  752 $ & $  427 $ & $  432 $ & $  811$ & $  813$
\\
$\text{branching ratio}$	& $0.8163$ & $0.0849$ & $0.0895$ & $0.0093$ & $0.902$ & $0.098$
\end{tabular}
\end{center}
\caption{ \label{tab:src}
Energy ($E$) and branching ratio
of the neutrino lines produced in the electron-capture decays of
${}^{51}\text{Cr}$ and ${}^{37}\text{Ar}$.
}
\end{table}

The reactor antineutrino anomaly can be explained by neutrino oscillations
with an oscillation length which is shorter than about 20 m.
Since the relation between a squared-mass difference $\Delta{m}^2$ and the corresponding oscillation length $L^{\text{osc}}$ is
\begin{equation}
L^{\text{osc}}
=
\frac{ 4 \pi E }{ \Delta{m}^{2} }
\simeq
2.5 \,
\frac{E \, [\text{MeV}]}{\Delta{m}^{2} \, [\text{eV}^{2}]}
\,
\text{m}
,
\label{Losc}
\end{equation}
and the average energy of the antineutrinos detected in a reactor experiment is about 4 MeV,
the required squared-mass differences is
\begin{equation}
\Delta{m}^2_{\text{SBL}}
\gtrsim
0.5 \, \text{eV}^2
.
\label{dm2rea}
\end{equation}

\subsection{The Gallium neutrino anomaly}
\label{sub:gallium}

The Gallium neutrino anomaly
\cite{Abdurashitov:2005tb,Laveder:2007zz,Giunti:2006bj,Acero:2007su,Giunti:2009zz,Giunti:2010zu,Giunti:2012tn},
is a short-baseline disappearance of $\nu_{e}$
measured in the
Gallium radioactive source experiments
GALLEX
\cite{Anselmann:1994ar,Hampel:1997fc,Kaether:2010ag}
and
SAGE
\cite{Abdurashitov:1996dp,Abdurashitov:1998ne,Abdurashitov:2005tb,Abdurashitov:2009tn}.

\begin{figure}
\begin{center}
\begin{minipage}[r]{0.49\linewidth}
\begin{center}
\subfigure[]{\label{fig:GaGe}
\includegraphics*[width=\linewidth]{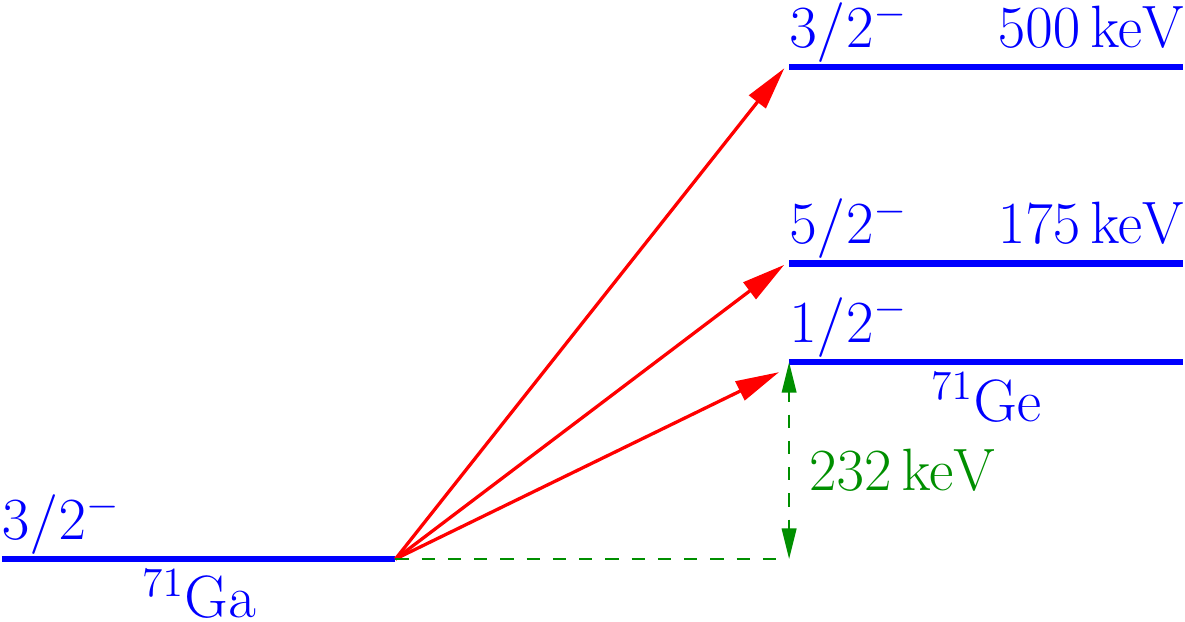}
}
\end{center}
\end{minipage}
\hfill
\begin{minipage}[l]{0.49\linewidth}
\begin{center}
\subfigure[]{\label{fig:gal}
\includegraphics*[width=0.8\linewidth]{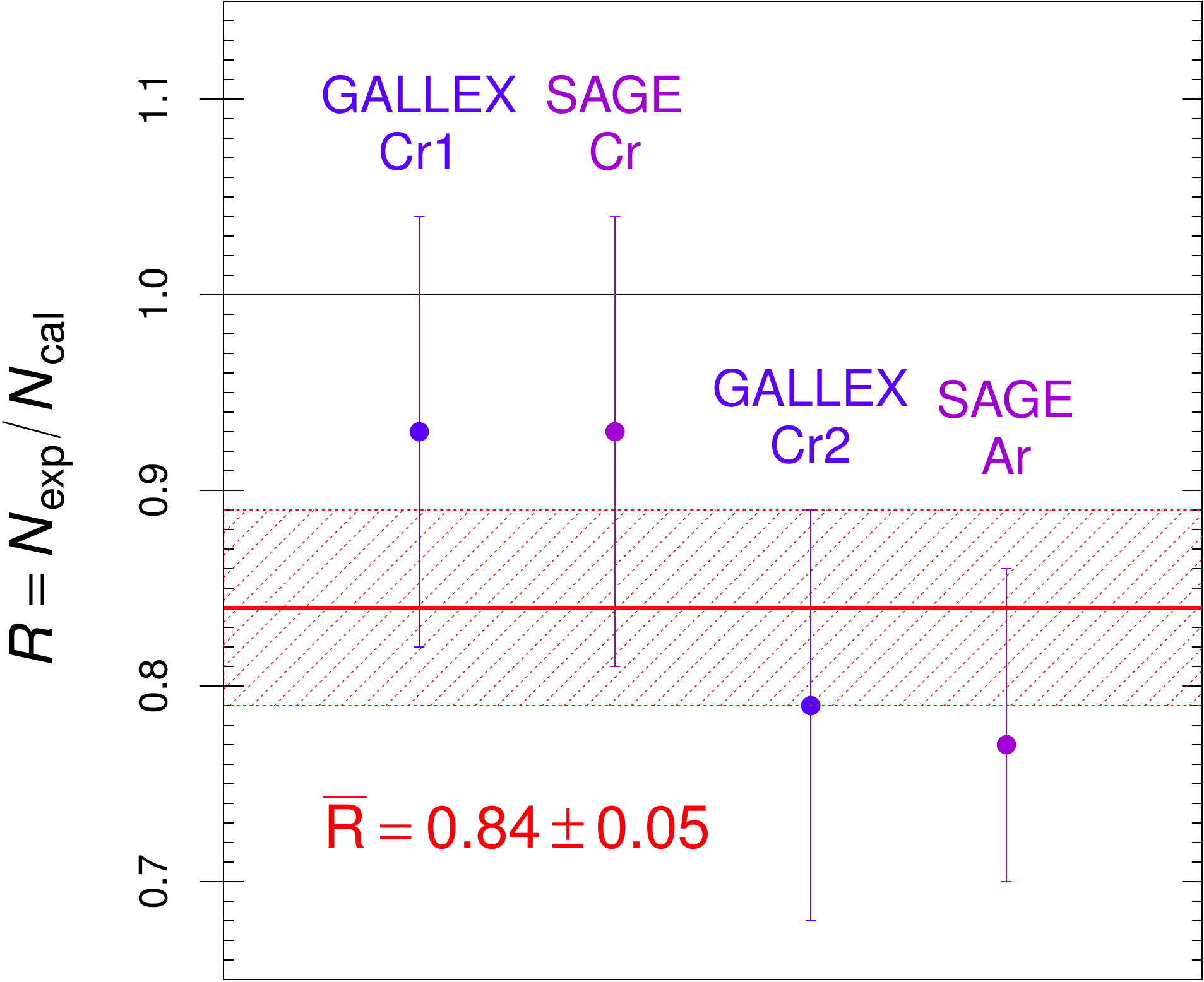}
}
\end{center}
\end{minipage}
\end{center}
\caption{ \label{fig:3.2}
\subref{fig:GaGe}:
${}^{71}\text{Ga}\to{}^{71}\text{Ge}$
transitions induced by
${}^{51}\text{Cr}$
and
${}^{37}\text{Ar}$
electron neutrinos.
\subref{fig:gal}:
Ratios $R$ of the measured
($N_{\text{exp}}$)
and calculated
($N_{\text{cal}}$)
number of electron neutrino events in
the GALLEX and SAGE
radioactive source experiments.
The horizontal shadowed red band shows the average ratio $\overline{R}$ and its uncertainty.
For each experiment
the error bar shows the experimental uncertainty.
}
\end{figure}

In these
radioactive source experiments
the detectors of the GALLEX and SAGE
solar neutrino experiments have been tested with
intense artificial ${}^{51}\text{Cr}$ and ${}^{37}\text{Ar}$ radioactive sources
which produce electron neutrinos through the electron captures
\begin{equation}
e^{-} + {}^{51}\text{Cr} \to {}^{51}\text{V} + \nu_{e}
,
\qquad
e^{-} + {}^{37}\text{Ar} \to {}^{37}\text{Cl} + \nu_{e}
.
\label{EC}
\end{equation}
The neutrino energies and branching ratios are given in
Tab.~\ref{tab:src}.
In each experiment
the radioactive source was placed near the center of the detector
and electron neutrinos have been detected with the reaction
\begin{equation}
\nu_{e} + {}^{71}\text{Ga} \to {}^{71}\text{Ge} + e^{-}
.
\label{f31}
\end{equation}
Figure~\ref{fig:GaGe}
shows the transitions
from the ground state of ${}^{71}\text{Ga}$ to the
ground state or one of two excited states of ${}^{71}\text{Ge}$
which are energetically allowed in the detection process.
The total detection cross section is given by
\begin{equation}
\sigma
=
\sigma_{\text{gs}}
\left(
1
+
\xi_{175}
\frac{\text{BGT}_{175}}{\text{BGT}_{\text{gs}}}
+
\xi_{500}
\frac{\text{BGT}_{500}}{\text{BGT}_{\text{gs}}}
\right)
,
\label{cs01}
\end{equation}
where
$\sigma_{\text{gs}}$
is the cross sections of the transitions
from the ground state of ${}^{71}\text{Ga}$ to the ground state of ${}^{71}\text{Ge}$,
$\text{BGT}_{\text{gs}}$
is the corresponding Gamow-Teller strength,
and
$\text{BGT}_{175}$
and
$\text{BGT}_{500}$
are the Gamow-Teller strengths of the transitions
from the ground state of ${}^{71}\text{Ga}$ to the two excited states of ${}^{71}\text{Ge}$
at about 175 keV and 500 keV (see Fig.~\ref{fig:GaGe}).
The coefficients of
$\text{BGT}_{175}/\text{BGT}_{\text{gs}}$
and
$\text{BGT}_{500}/\text{BGT}_{\text{gs}}$
are determined by phase space:
$\xi_{175}({}^{51}\text{Cr}) = 0.669$,
$\xi_{500}({}^{51}\text{Cr}) = 0.220$,
$\xi_{175}({}^{37}\text{Ar}) = 0.695$,
$\xi_{500}({}^{37}\text{Ar}) = 0.263$
\cite{Bahcall:1997eg}.

The cross sections of the transitions
from the ground state of ${}^{71}\text{Ga}$ to the ground state of ${}^{71}\text{Ge}$
have been calculated accurately by Bahcall
\cite{Bahcall:1997eg}:
\begin{equation}
\sigma_{\text{gs}}({}^{51}\text{Cr})
=
55.3 \times 10^{-46} \, \text{cm}^2
,
\qquad
\sigma_{\text{gs}}({}^{37}\text{Ar})
=
66.2 \times 10^{-46} \, \text{cm}^2
,
\label{sgs}
\end{equation}
and
\cite{Haxton:1998uc,Giunti:2012tn}
\begin{equation}
\text{BGT}_{\text{gs}}
=
0.0871 \pm 0.0004
.
\label{f32}
\end{equation}
The Gamow-Teller strengths
$\text{BGT}_{175}$
and
$\text{BGT}_{500}$
have been measured in 1985 in the
$(p,n)$
experiment of Krofcheck et al. \cite{Krofcheck:1985fg,Krofcheck-PhD-1987}
(see Table I of Ref.~\cite{Bahcall:1997eg})
and,
more precisely,
in 2011 in the
$({}^{3}\text{He},{}^{3}\text{H})$
experiment of Frekers et al.
\cite{Frekers:2011zz}.
The results are listed in Tab.~\ref{tab:bgt}.

Figure~\ref{fig:gal} shows the
ratios $R$ of the number of electron neutrino events
($N_{\text{exp}}$)
measured in the GALLEX and SAGE
radioactive source experiments
and that calculated
($N_{\text{cal}}$)
with the Frekers et al. Gamow-Teller strengths.
The average ratio shown in the figure is
$\overline{R} = 0.84 \pm 0.05$,
which indicates a deficit with a nominal
statistical significance of about
$2.9\sigma$.

Since the average neutrino traveling distances
in the GALLEX and SAGE
radioactive source experiments
are
$\langle L \rangle_{\text{GALLEX}} = 1.9 \, \text{m}$
and
$\langle L \rangle_{\text{SAGE}} = 0.6 \, \text{m}$,
from Eq.~(\ref{Losc}) and Tab.~\ref{tab:src}
one can estimate that
the Gallium neutrino anomaly can be explained by neutrino oscillations
generated by a squared-mass difference
\begin{equation}
\Delta{m}^2_{\text{SBL}}
\gtrsim
1 \, \text{eV}^2
.
\label{dm2gal}
\end{equation}

\begin{table}[t]
\begin{center}
\begin{tabular}{lccc}
Reference
&
Method
&
$\text{BGT}_{175}$
&
$\text{BGT}_{500}$
\\
\hline
Krofcheck et al.
\protect\cite{Krofcheck:1985fg,Krofcheck-PhD-1987}
&
${}^{71}\text{Ga} (p,n) {}^{71}\text{Ge}$
&
$< 0.005$
&
$
0.011
\pm
0.002
$
\\
Frekers et al.
\protect\cite{Frekers:2011zz}
&
${}^{71}\text{Ga} ({}^{3}\text{He},{}^{3}\text{H}) {}^{71}\text{Ge}$
&
$
0.0034
\pm
0.0026
$
&
$
0.0176
\pm
0.0014
$
\end{tabular}
\end{center}
\caption{
\label{tab:bgt}
Values of the Gamow-Teller strengths of the transitions
from the ground state of ${}^{71}\text{Ga}$ to the two excited states of ${}^{71}\text{Ge}$
at 175 keV and 500 keV.
}
\end{table}

\subsection{The LSND anomaly}
\label{sub:LSND}

The LSND experiment
\cite{Athanassopoulos:1995iw,Aguilar:2001ty}
observed an excess of $\bar\nu_{e}$ events
in a beam of $\bar\nu_{\mu}$ produced by $\mu^{+}$
decay at rest,
\begin{equation}
\mu^{+} \to e^{+} + \nu_{e} + \bar\nu_{\mu}
.
\label{l008}
\end{equation}
The energy spectrum of $\bar\nu_{\mu}$ is
$
\phi_{\bar\nu_{\mu}}(E)
\propto
E^2 \left( 3 - 4 E / m_{\mu} \right)
$
(see Ref.~\cite{Fukugita:2003en})
for neutrino energies $E$ smaller than
\begin{equation}
E_{\text{max}} = (m_{\mu}-m_{e})/2 \simeq 52.6 \, \text{MeV}
.
\label{l010}
\end{equation}
The $\bar\nu_{e}$ events
have been detected at a distance $L \simeq 30 \, \text{m}$
through the inverse neutron decay process (\ref{l011})
in a detector filled with liquid scintillator
in the range
$20 \lesssim E_{e} \lesssim 60 \, \text{MeV}$
for the energy $E_{e}$ of the detected positron.

The nominal statistical significance
of the LSND $\bar\nu_{e}$ appearance signal
is of about $3.8\sigma$.
However,
it must be noted that the similar KARMEN experiment
\cite{Armbruster:2000uw,Armbruster:2002mp}
did not measure any excess of $\bar\nu_{e}$ events
over the background at a distance $L \simeq 18 \, \text{m}$.

From Eq.~(\ref{Losc})
one can estimate that
the LSND $\bar\nu_{e}$ appearance signal
can be explained by $\bar\nu_{\mu}\to\bar\nu_{e}$ oscillations
generated by a squared-mass difference
\begin{equation}
\Delta{m}^2_{\text{SBL}}
\gtrsim
0.1 \, \text{eV}^2
.
\label{dm2lsnd}
\end{equation}
\begin{table}[t]
\begin{center}
\begin{tabular}{c|cccc|cc}
					&3+1									&3+1									&3+1									&3+1									&3+2									&3+2\\
					&GLO									&PrGLO									&noMB									&noLSND									&GLO									&PrGLO\\
\hline
$\chi^{2}_{\text{min}}$			&306.0			&276.3			&251.2			&291.3			&299.6			&271.1			\\
NDF					&268			&262			&230			&264			&264			&258			\\
GoF					& 5\%		&26\%		&16\%		&12\%		& 7\%		&28\%		\\
\hline
$(\chi^{2}_{\text{min}})_{\text{APP}}$	&98.9			&77.0			&50.9			&91.8			&86.0			&69.6			\\
$(\chi^{2}_{\text{min}})_{\text{DIS}}$	&194.4			&194.4			&194.4			&194.4			&192.9			&192.9			\\
	$\Delta\chi^{2}_{\text{PG}}$	&13.0		&5.3		&6.2		&5.3		&20.7		&8.6		\\
	$\text{NDF}_{\text{PG}}$	&2		&2		&2		&2		&4		&4		\\
	$\text{GoF}_{\text{PG}}$	&0.1\%	& 7\%	& 5\%	& 7\%	&0.04\%	& 7\%	\\
\hline
$\Delta\chi^{2}_{\text{NO}}$		&$49.2$		&$47.7$		&$48.1$		&$11.4$		&$55.7$		&$52.9$		\\
	$\text{NDF}_{\text{NO}}$	&$3$		&$3$		&$3$		&$3$		&$7$		&$7$		\\
	$n\sigma_{\text{NO}}$		&$6.4\sigma$	&$6.3\sigma$	&$6.4\sigma$	&$2.6\sigma$	&$6.1\sigma$	&$5.9\sigma$	\\
\end{tabular}
\end{center}
\caption{ \label{tab:chi}
Results of the fit of short-baseline data
taking into account all MiniBooNE data (GLO),
only the MiniBooNE data above 475 MeV (PrGLO),
without MiniBooNE data (noMB)
and without LSND data (noLSND)
in the
3+1 and 3+2 schemes.
The first three lines give
the minimum $\chi^{2}$ ($\chi^{2}_{\text{min}}$),
the number of degrees of freedom (NDF) and
the goodness-of-fit (GoF).
The following five lines give the quantities
relevant for the appearance-disappearance (APP-DIS) parameter goodness-of-fit (PG)
\protect\cite{Maltoni:2003cu}.
The last three lines give
the difference between the $\chi^{2}$ without short-baseline oscillations and $\chi^{2}_{\text{min}}$
($\Delta\chi^{2}_{\text{NO}}$),
the corresponding difference of number of degrees of freedom ($\text{NDF}_{\text{NO}}$)
and the resulting
number of $\sigma$'s ($n\sigma_{\text{NO}}$) for which the absence of oscillations is disfavored.
}
\end{table}

\section{Global fits of short-baseline data}
\label{sec:global}

Many analyses of
short-baseline neutrino oscillation data
have been done since the discovery of the LSND anomaly in the middle 90's
\cite{GomezCadenas:1995sj,Goswami:1995yq,Okada:1996kw,Bilenky:1996rw,Bilenky:1999ny,Giunti:2000wt,Barger:2000ch,Peres:2000ic,Grimus:2001mn,Strumia:2002fw,Maltoni:2002xd,Foot:2002tf,Giunti:2003cf,Sorel:2003hf,Barger:2003xm,Maltoni:2007zf,Goswami:2007kv,Schwetz:2007cd,Karagiorgi:2009nb,Akhmedov:2010vy,Nelson:2010hz}.
The activity became more interesting after the discoveries of the Gallium neutrino anomaly
\cite{Giunti:2006bj,Acero:2007su,Giunti:2007xv,Giunti:2009zz,Giunti:2010wz,Giunti:2010zu,Giunti:2012tn,Giunti:2010zs,Giunti:2010jt,Giunti:2010uj}
and the reactor antineutrino anomaly
\cite{Mention:2011rk,Kopp:2011qd,Donini:2011jh,Conrad:2011ce,Giunti:2011gz,Giunti:2011hn,Giunti:2011cp,Karagiorgi:2012kw,Kuflik:2012sw,Donini:2012tt,Archidiacono:2012ri,Giunti:2012tn,Giunti:2012bc,Conrad:2012qt}.
The last updated global fits of short-baseline neutrino oscillation data have been presented in
Refs.~\cite{Kopp:2013vaa,Giunti:2013aea}.
These analyses take into account the final results of the
MiniBooNE experiment,
which was made in order to check the LSND signal
with about one order of magnitude larger distance ($L$) and energy ($E$),
but the same order of magnitude for the ratio $L/E$
from which neutrino oscillations depend.
Unfortunately, the results of the
MiniBooNE experiment are ambiguous,
because the LSND signal was not seen in the neutrino mode
($\nu_{\mu}\to\nu_{e}$)
\cite{AguilarArevalo:2008rc}
and the $\bar\nu_{\mu}\to\bar\nu_{e}$ signal observed in 2010 \cite{AguilarArevalo:2010wv}
with the first half of the antineutrino data
was not observed in the second half of the antineutrino data
\cite{Aguilar-Arevalo:2013pmq}.
Moreover,
the MiniBooNE data in both neutrino and antineutrino modes
show an excess in the low-energy bins
which is widely considered to be anomalous
because it is at odds with neutrino oscillations
\cite{Giunti:2011hn,Giunti:2011cp}\footnote{
The interesting possibility of reconciling the low-energy anomalous data with neutrino oscillations
through energy reconstruction effects proposed in Ref.~\cite{Martini:2012fa,Martini:2012uc}
still needs a detailed study.
}.

In the following we discuss the results of an update\footnote{
We added the data of several reactor neutrino experiments:
the old results of
Rovno88 \cite{Afonin:1988gx}
and
SRP \cite{Greenwood:1996pb},
which were omitted in our previous analyses
\cite{Giunti:2012tn,Giunti:2013aea}
but have been considered in the fits of other authors
\cite{Mention:2011rk,Kopp:2013vaa};
the old results of
Chooz \cite{Apollonio:2002gd}
and
Palo Verde \cite{Boehm:2001ik}
as suggested in Ref.\cite{Zhang:2013ela};
the new results of
Double Chooz \cite{Abe:2014bwa}
and
Daya Bay \cite{Zhang:2015fya,Zhan:2015aha,An:2015nua}.
}
of the global fit of short-baseline neutrino oscillation
data presented in Ref.~\cite{Giunti:2013aea}
in which the data of the following three groups of experiments
are considered:

\begin{enumerate}

\renewcommand{\labelenumi}{(\theenumi)}
\renewcommand{\theenumi}{\Alph{enumi}}

\item
The
$\nua{\mu}\to\nua{e}$
appearance data of the
LSND \cite{Aguilar:2001ty},
MiniBooNE \cite{Aguilar-Arevalo:2013pmq},
BNL-E776 \cite{Borodovsky:1992pn},
KARMEN \cite{Armbruster:2002mp},
NOMAD \cite{Astier:2003gs},
ICARUS \cite{Antonello:2013gut}
and
OPERA \cite{Agafonova:2013xsk}
experiments\footnote{
For simplicity we do not take into account the correct analysis of
the ICARUS and OPERA data
presented in Ref.~\cite{Palazzo:2015wea}
(see also Ref.~\cite{Donini:2007yf})
because it would not change significantly the results of the global fits.
}.

\item
The following
$\nua{e}$
disappearance data:
1)
the data of the
Bugey-4 \cite{Declais:1994ma},
ROVNO91 \cite{Kuvshinnikov:1990ry},
Bugey-3 \cite{Declais:1995su},
Gosgen \cite{Zacek:1986cu},
ILL \cite{Hoummada:1995zz},
Krasnoyarsk \cite{Vidyakin:1990iz},
Rovno88 \cite{Afonin:1988gx},
SRP \cite{Greenwood:1996pb},
Chooz \cite{Apollonio:2002gd},
Palo Verde \cite{Boehm:2001ik},
Double Chooz \cite{Abe:2014bwa}, and
Daya Bay \cite{Zhang:2015fya,Zhan:2015aha,An:2015nua}
reactor antineutrino experiments,
with the new theoretical fluxes
\cite{Mueller:2011nm,Huber:2011wv,Mention:2011rk,Abazajian:2012ys}
(see Section~\ref{sub:reactor});
2)
the data of the
GALLEX
\cite{Anselmann:1994ar,Hampel:1997fc,Kaether:2010ag}
and
SAGE
\cite{Abdurashitov:1996dp,Abdurashitov:1998ne,Abdurashitov:2005tb,Abdurashitov:2009tn}
Gallium radioactive source experiments
with the statistical method discussed in Ref.~\cite{Giunti:2010zu},
considering the recent
${}^{71}\text{Ga}({}^{3}\text{He},{}^{3}\text{H}){}^{71}\text{Ge}$
cross section measurement in Ref.~\cite{Frekers:2011zz}
(see Section~\ref{sub:gallium});
3)
the solar neutrino constraint on $\sin^{2}2\vartheta_{ee}$
\cite{Giunti:2009xz,Palazzo:2011rj,Palazzo:2012yf,Giunti:2012tn,Palazzo:2013me};
4)
the
KARMEN \cite{Bodmann:1994py,Armbruster:1998uk}
and
LSND \cite{Auerbach:2001hz}
$\nu_{e} + {}^{12}\text{C} \to {}^{12}\text{N}_{\text{g.s.}} + e^{-}$
scattering data \cite{Conrad:2011ce},
with the method discussed in Ref.~\cite{Giunti:2011cp}.

\item
The constraints on
$\nua{\mu}$
disappearance obtained from
the data of the
CDHSW experiment \cite{Dydak:1983zq},
from the analysis \cite{Maltoni:2007zf} of
the data of
atmospheric neutrino oscillation experiments\footnote{
The analysis of the IceCube data
\cite{Barger:2011rc,Razzaque:2012tp,Esmaili:2012nz,Esmaili:2013vza},
which could give a marginal contribution,
have not been considered
because it is too complicated and subject to large uncertainties.
},
from the analysis \cite{Hernandez:2011rs,Giunti:2011hn} of the
MINOS neutral-current data \cite{Adamson:2011ku}
and from the analysis of the
SciBooNE-MiniBooNE
neutrino \cite{Mahn:2011ea} and antineutrino \cite{Cheng:2012yy} data.

\end{enumerate}

\begin{figure*}[t]
\null
\hfill
\includegraphics*[width=0.49\linewidth]{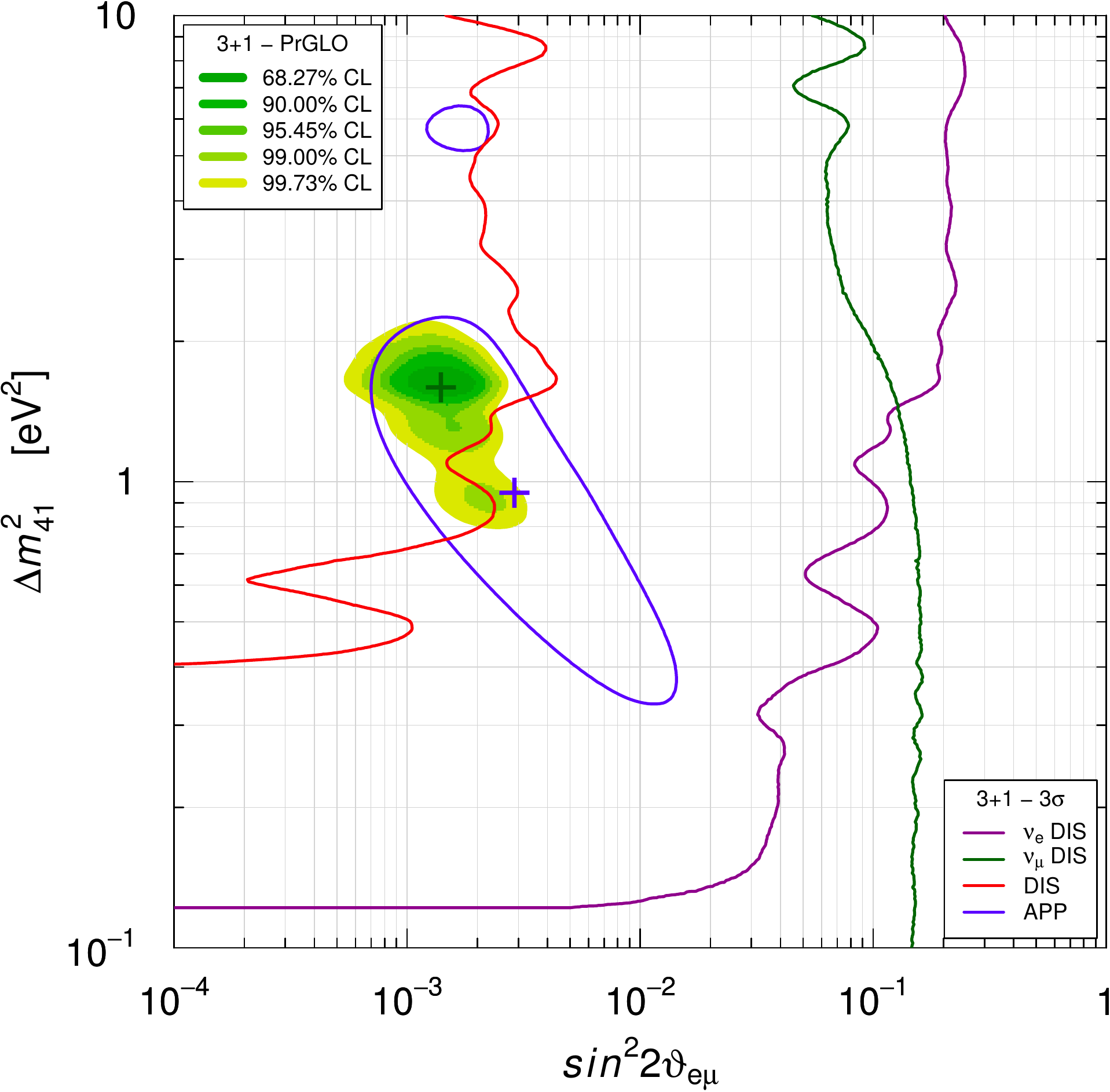}
\hfill
\includegraphics*[width=0.49\linewidth]{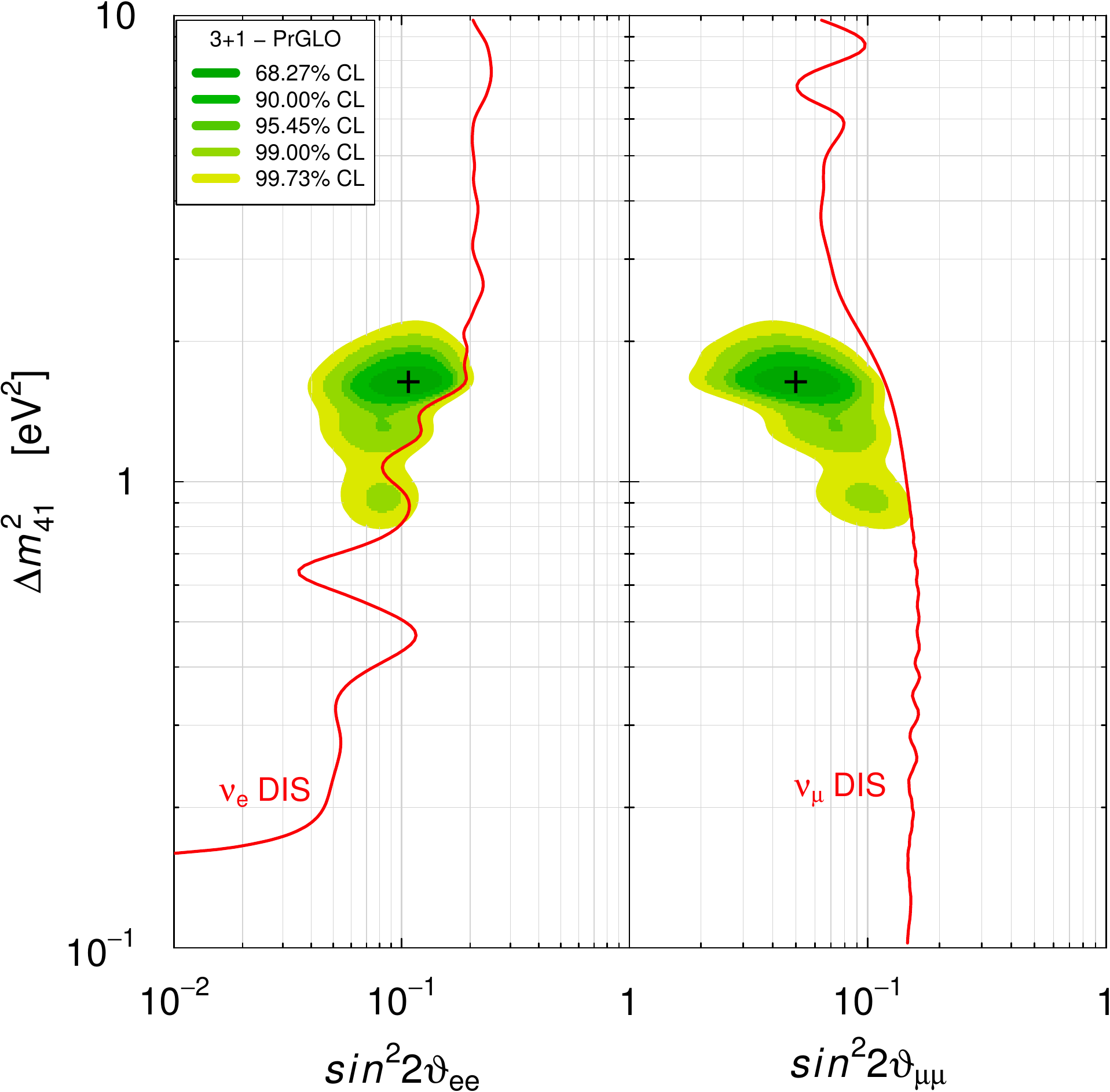}
\hfill
\null
\caption{ \label{fig:glo}
Allowed regions in the
$\sin^{2}2\vartheta_{e\mu}$--$\Delta{m}^{2}_{41}$,
$\sin^{2}2\vartheta_{ee}$--$\Delta{m}^{2}_{41}$
and
$\sin^{2}2\vartheta_{\mu\mu}$--$\Delta{m}^{2}_{41}$
planes
obtained in the pragmatic 3+1-PrGLO global fit
of short-baseline neutrino oscillation data
compared with the $3\sigma$ allowed regions
obtained from
$\protect\nua{\mu}\to\protect\nua{e}$
short-baseline appearance data (APP)
and the $3\sigma$ constraints obtained from
$\protect\nua{e}$
short-baseline disappearance data ($\nu_{e}$ DIS),
$\protect\nua{\mu}$
short-baseline disappearance data ($\nu_{\mu}$ DIS)
and the
combined short-baseline disappearance data (DIS).
The best-fit points of the PrGLO and APP fits are indicated by crosses.
}
\end{figure*}

Table~\ref{tab:chi}
summarizes the statistical results obtained
from global fits of the data above
in the
3+1 and 3+2 schemes
(see Section~\ref{sec:mixing};
for a global fit in the 3+1+1 scheme see Ref.~\cite{Giunti:2013aea}).
In the global (GLO) fits
all the MiniBooNE data are considered,
including the anomalous low-energy bins,
which are omitted in the pragmatic PrGLO global fits \cite{Giunti:2013aea}.
There is also a 3+1-noMB fit without MiniBooNE data
and
a 3+1-noLSND fit without LSND data.

From Tab.\ref{tab:chi},
one can see that in all fits which include the LSND data
the absence of short-baseline oscillations
is nominally disfavored by about $6\sigma$,
because the improvement of the $\chi^2$ with short-baseline oscillations
is much larger than the number of oscillation parameters.

In all the
3+1 and 3+2 schemes
the goodness-of-fit in the GLO analysis is significantly worse than that in the PrGLO analysis
and the appearance-disappearance parameter goodness-of-fit is much worse.
This result confirms the fact that the MiniBooNE low-energy anomaly
is incompatible with neutrino oscillations,
because it would require a small value of $\Delta{m}^2_{41}$
and a large value of $\sin^22\vartheta_{e\mu}$
\cite{Giunti:2011hn,Giunti:2011cp},
which are excluded by the data of other experiments
(see Ref.~\cite{Giunti:2013aea} for further details)\footnote{
One could fit the three anomalous MiniBooNE low-energy bins
in a 3+2 scheme \cite{Conrad:2012qt}
by considering the appearance data without the
ICARUS \cite{Antonello:2013gut}
and
OPERA \cite{Agafonova:2013xsk}
constraints,
but the required large transition probability is excluded
by the disappearance data.
}.
Note that the appearance-disappearance tension
in the 3+2-GLO fit is even worse than that in the 3+1-GLO fit,
since the $\Delta\chi^{2}_{\text{PG}}$ is so much larger that it cannot be compensated
by the additional degrees of freedom\footnote{
This behavior has been explained in Ref.~\cite{Archidiacono:2013xxa}.
It was found also in the analysis presented in Ref.~\cite{Kopp:2013vaa}.
}.
Therefore,
we think that it is very likely that the MiniBooNE low-energy anomaly
has an explanation which is different from neutrino oscillations.
The cause of the MiniBooNE low-energy excess of $\nu_{e}$-like events
is going to be investigated in the MicroBooNE experiment at Fermilab
\cite{MicroBooNE-2007,Szelc:2015dga},
which is a large Liquid Argon Time Projection Chamber (LArTPC)
in which electrons and photons can be distinguished\footnote{
In the MiniBooNE  mineral-oil Cherenkov detector
$\nu_{e}$-induced events
cannot be distinguished from
$\nu_{\mu}$-induced events which produce only a visible photon
(for example neutral-current $\pi^{0}$ production in which only one of the two decay photons is visible).
}
(see the review in Ref.~\cite{Katori:2014qta}).

In the following we adopt the ``pragmatic approach'' advocated in Ref.~\cite{Giunti:2013aea}
which considers the PrGLO fits,
without the anomalous MiniBooNE low-energy bins,
as more reliable than the GLO fits,
which include the anomalous MiniBooNE low-energy bins.

The 3+2 mixing scheme
was considered to be interesting in 2010
when the MiniBooNE neutrino
\cite{AguilarArevalo:2008rc}
and antineutrino
\cite{AguilarArevalo:2010wv}
data showed a CP-violating tension,
but
this tension almost disappeared in the final MiniBooNE data
\cite{Aguilar-Arevalo:2013pmq}.
In fact, from Tab.\ref{tab:chi}
one can see that there is little improvement of the 3+2-PrGLO fit
with respect to the 3+1-PrGLO fit,
in spite of the four additional parameters and the additional possibility of CP violation.
Moreover,
since the p-value obtained by restricting the 3+2 scheme to 3+1
disfavors the 3+1 scheme only at
$1.1\sigma$,
we think that considering the larger complexity of the 3+2 scheme
is not justified by the data\footnote{
See however the alternative discussion in Ref.~\cite{Kopp:2013vaa}.
}.

Figure~\ref{fig:glo}
shows the allowed regions in the
$\sin^{2}2\vartheta_{e\mu}$--$\Delta{m}^{2}_{41}$,
$\sin^{2}2\vartheta_{ee}$--$\Delta{m}^{2}_{41}$ and
$\sin^{2}2\vartheta_{\mu\mu}$--$\Delta{m}^{2}_{41}$
planes
obtained in the 3+1-PrGLO fit.
These regions are relevant, respectively, for
$\nua{\mu}\to\nua{e}$ appearance,
$\nua{e}$ disappearance and
$\nua{\mu}$ disappearance
searches.
The corresponding marginal allowed intervals of the oscillation parameters
are given in Tab.\ref{tab:int}.
Figure~\ref{fig:glo}
shows also the region allowed by $\nua{\mu}\to\nua{e}$ appearance data
and
the constraints from
$\nua{e}$ disappearance and
$\nua{\mu}$ disappearance data.
One can see that the combined disappearance constraint
in the $\sin^{2}2\vartheta_{e\mu}$--$\Delta{m}^{2}_{41}$ plane
excludes a large part of the region allowed by $\nua{\mu}\to\nua{e}$ appearance data,
leading to the well-known
appearance-disappearance tension
\cite{Kopp:2011qd,Giunti:2011gz,Giunti:2011hn,Giunti:2011cp,Conrad:2012qt,Archidiacono:2012ri,Archidiacono:2013xxa,Kopp:2013vaa}
quantified by the parameter goodness-of-fit in Tab.\ref{tab:chi}.

\begin{table}[t]
\begin{center}
\begin{tabular}{c|cccc}
CL
&
$\Delta{m}^2_{41}[\text{eV}^2]$
&
$\sin^22\vartheta_{e\mu}$
&
$\sin^22\vartheta_{ee}$
&
$\sin^22\vartheta_{\mu\mu}$
\\
\hline
68.27\%
&
$ 1.57 - 1.72 $
&
$ 0.0011 - 0.0018 $
&
$ 0.085 - 0.13 $
&
$ 0.039 - 0.066 $
\\
\hline
90.00\%
&
$ 1.53 - 1.78 $
&
$ 0.00098 - 0.0020 $
&
$ 0.071 - 0.15 $
&
$ 0.032 - 0.078 $
\\
\hline
95.45\%
&
$ 1.50 - 1.84 $
&
$ 0.00089 - 0.0021 $
&
$ 0.063 - 0.16 $
&
$ 0.030 - 0.085 $
\\
\hline
99.00\%
&
$ 1.24 - 1.95 $
&
$ 0.00074 - 0.0023 $
&
$ 0.054 - 0.18 $
&
$ 0.025 - 0.095 $
\\
\hline
99.73\%
&
$ 0.87 - 2.04 $
&
$ 0.00065 - 0.0026 $
&
$ 0.046 - 0.19 $
&
$ 0.021 - 0.12 $
\end{tabular}
\end{center}
\caption{ \label{tab:int}
Marginal allowed intervals of the oscillation parameters
obtained in the global 3+1-PrGLO fit
of short-baseline neutrino oscillation data.
}
\end{table}

It is interesting to investigate what is the
impact of the MiniBooNE experiment
on the global analysis of short-baseline neutrino oscillation data.
With this aim,
we consider two additional 3+1 fits:
a 3+1-noMB fit without MiniBooNE data
and
a 3+1-noLSND fit without LSND data.
From Tab.\ref{tab:chi}
one can see that the results of the
3+1-noMB fit are similar to those of the
3+1-PrGLO fit
and the nominal exclusion of the case of no-oscillations remains at the level of $6\sigma$.
On the other hand,
in the 3+1-noLSND fit,
without LSND data,
the nominal exclusion of the case of no-oscillations drops dramatically to
$2.6\sigma$.
In fact,
in this case
the main indication in favor of short-baseline oscillations
is given by the reactor
and
Gallium
anomalies
which have a similar statistical significance
\cite{Giunti:2012tn}.
Therefore,
it is clear that the LSND experiment is still crucial for the indication in favor of short-baseline
$\bar\nu_{\mu}\to\bar\nu_{e}$
transitions
and the MiniBooNE experiment has been rather inconclusive.
\section{$\beta$ decay and neutrinoless double-$\beta$ decay}
\label{sec:decay}

The existence of massive neutrinos at the eV scale
can be probed in
$\beta$-decay experiments
\cite{Kraus:2012he,Belesev:2012hx,Belesev:2013cba,Riis:2010zm,Formaggio:2011jg,SejersenRiis:2011sj,Esmaili:2012vg}
and in
neutrinoless double-$\beta$ decay experiments
\cite{Goswami:2005ng,Goswami:2007kv,Barry:2011wb,Li:2011ss,Rodejohann:2012xd,Giunti:2012tn,Girardi:2013zra,Pascoli:2013fiz,Meroni:2014tba,Abada:2014nwa,Giunti:2015kza,Pas:2015eia}.

\subsection{Tritium $\beta$ decay}
\label{sec:beta}

The most sensitive experiments on the search of the effects of neutrino masses in $\beta$ decay
use the Tritium decay process
\begin{equation}
{}^{3}\text{H} \to
{}^{3}\text{He} + e^{-} + \bar\nu_{e}
.
\label{tritium}
\end{equation}
Non-zero neutrino masses distort the measurable spectrum of the emitted electron.
It is convenient to consider the Kurie function (see, for example, Ref.~\cite{Giunti:2007ry})
\begin{equation}
K^2(T)
=
\left( Q - T_{e} \right)
\sum_{k}
|U_{ek}|^2
\sqrt{ \left( Q - T_{e} \right)^{2} - m_{k}^{2} }
\,
\Theta(Q-T_{e}-m_{k})
,
\label{kurie1}
\end{equation}
where $T_{e}$ is the electron kinetic energy,
$
Q
=
M_{{}^{3}\text{H}}
-
M_{{}^{3}\text{He}}
-
m_{e}
\simeq
18.574 \, \text{keV}
$
is the $Q$-value of the process,
and
$\Theta$ is the Heaviside step function.
Considering an experiment in which the energy resolution is such that
$m_{k} \ll Q-T_{e}$
for the three standard light neutrino masses ($k=1,2,3$),
the Kurie function can be approximated by
\begin{align}
K^2(T)
\simeq
\null & \null
\left( Q - T_{e} \right)
\sqrt{ \left( Q - T_{e} \right)^{2} - m_{\beta}^{2} }
\,
\Theta(Q-T_{e}-m_{\beta})
\nonumber
\\
\null & \null
+
\left( Q - T_{e} \right)
\sum_{k\geq4}
|U_{ek}|^2
\sqrt{ \left( Q - T_{e} \right)^{2} - m_{k}^{2} }
\,
\Theta(Q-T_{e}-m_{k})
,
\label{kurie2}
\end{align}
with the effective light neutrino mass
\begin{equation}
m_{\beta} = \left(\sum_{k=1}^{3} |U_{ek}|^2 m_{k}^2 \right)^{1/2}
.
\label{efnmass}
\end{equation}
The most stringent 95\% CL limits on $m_{\beta}$
have been obtained
in the Mainz \cite{Kraus:2004zw} and Troitsk \cite{Aseev:2011dq} experiments:
\begin{align}
\null & \null
m_{\beta} \leq 2.3 \, \text{eV}
\quad
(\text{95\% C.L., Mainz})
,
\label{B088}
\\
\null & \null
m_{\beta} \leq 2.1 \, \text{eV}
\quad
(\text{95\% C.L., Troitsk})
.
\label{B089}
\end{align}
The experiment KATRIN \cite{Mertens:2015ila},
which is under construction and is scheduled to start data taking in 2016,
will aim to reach a sensitivity of $0.2 \, \text{eV}$ at 90\% C.L. for $m_{\beta}$ in five years of running\footnote{
Other future projects are described in the reviews \cite{Dragoun:2015oja,Drexlin:2013lha,Formaggio:2014ppa}
}.

\begin{figure}
\begin{center}
\begin{minipage}[r]{0.46\linewidth}
\begin{center}
\subfigure[]{\label{fig:mainz}
\includegraphics*[width=\linewidth]{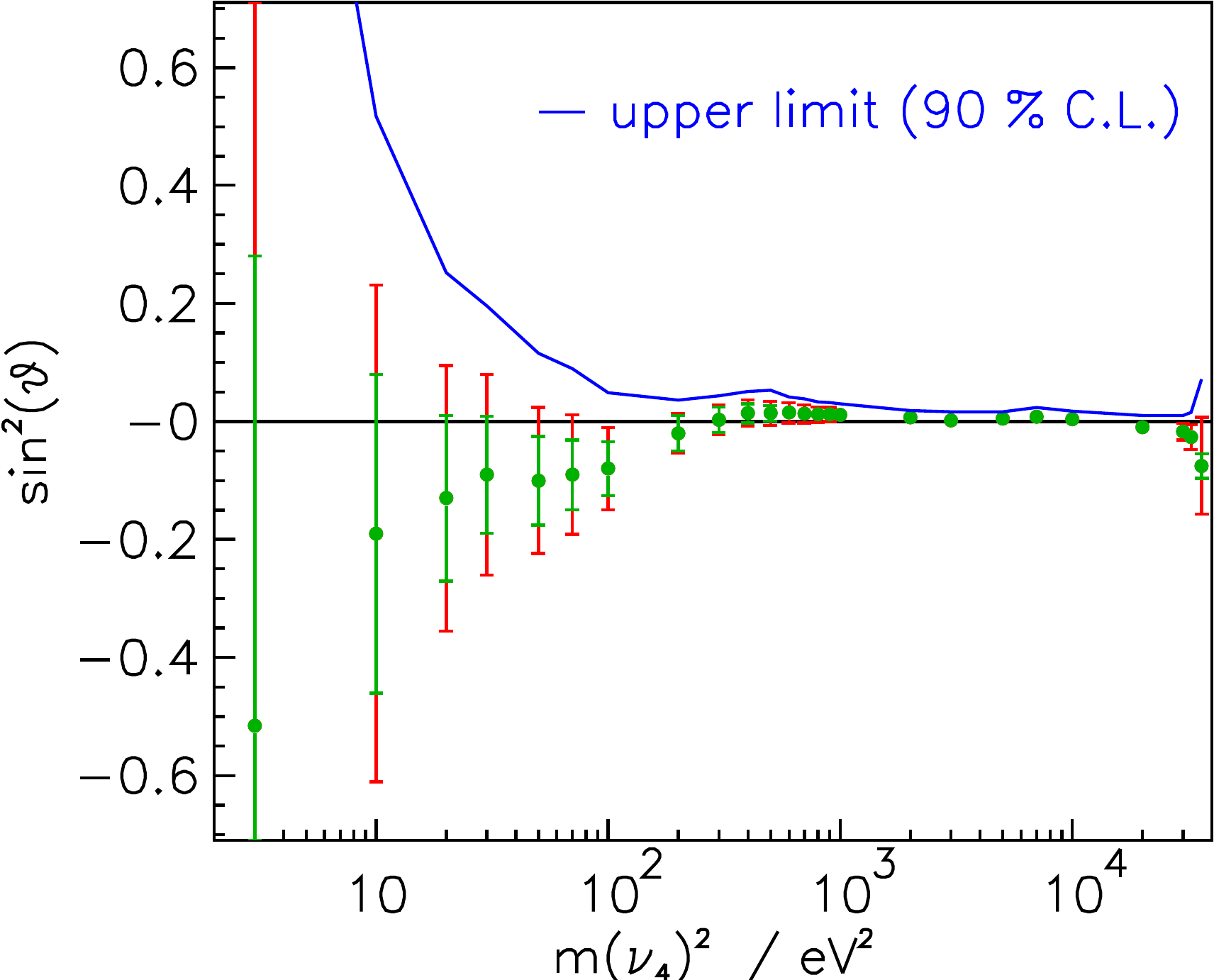}
}
\end{center}
\end{minipage}
\hfill
\begin{minipage}[l]{0.52\linewidth}
\begin{center}
\subfigure[]{\label{fig:troitsk}
\includegraphics*[width=\linewidth]{fig-06b.pdf}
}
\end{center}
\end{minipage}
\end{center}
\caption{ \label{fig:beta}
\subref{fig:mainz}:
Upper 90\% C.L. limit on $\sin^2(\vartheta)=|U_{e4}|^2$
as a function of $m(\nu_{4})^2=m_{4}^2$
obtained in the Mainz experiment
\protect\cite{Kraus:2012he}.
\subref{fig:troitsk}:
Upper 95\% C.L. limits on $U_{e4}^2=|U_{e4}|^2$
as functions of $m_{4}^2$
obtained with different statistical methods in the Troitsk experiment
\protect\cite{Belesev:2013cba}.
}
\end{figure}

The expression (\ref{kurie2}) of the Kurie function shows that
a heavy nonstandard neutrino mass $m_{k}$ with $k\geq4$ can be measured by observing
a kink of the kinetic energy spectrum of the emitted electron at
$Q-m_{k}$
below the end point
\cite{Shrock:1980vy,Schreckenbach:1983cg,Ohshima:1993pp,Mortara:1993iv,Farzan:2001cj,Farzan:2002zq,deGouvea:2006gz,Riis:2010zm,Giunti:2011cp}.
Recently,
the Mainz
\cite{Kraus:2012he}
and
Troitsk
\cite{Belesev:2012hx,Belesev:2013cba}
collaborations reanalyzed the measured spectra
which led to the limits (\ref{B088}) and (\ref{B089})
searching for the effect of a heavy neutrino mass $m_{4}$.
The negative results led to the upper bounds for the mixing factor $|U_{e4}|^2$
shown in Fig.~\ref{fig:beta}.
These bounds imply that the squared-mass difference that can explain the
reactor and Gallium anomalies
(see Sections~\ref{sub:reactor} and \ref{sub:gallium})
must be smaller than about $40 \, \text{eV}^2$
at 90\% C.L.
\cite{Giunti:2012bc}.
This is encouraging for future radioactive source experiments
which aim at a ``smoking-gun'' measurement of short-baseline oscillations
as a function of distance and/or energy,
because the predicted range for the oscillation length
$
L^{\text{osc}}_{41}
=
4 \pi E / \Delta{m}^{2}_{41}
$
is
\cite{Giunti:2012bc}
\begin{equation}
6 \, \text{cm}
\lesssim
\frac{L^{\text{osc}}_{41}}{E\,[\text{MeV}]}
\lesssim
3 \, \text{m}
\qquad
(2\sigma)
.
\label{L41}
\end{equation}

Some studies were performed to analyze the sensitivity of the KATRIN experiment
\cite{Mertens:2015ila} to the effects of
heavy sterile neutrinos with keV-scale masses \cite{Abdurashitov:2015jta,Mertens:2014nha,Barry:2014ika}
and
light eV-scale sterile neutrinos
\cite{Riis:2010zm,Formaggio:2011jg,SejersenRiis:2011sj,Esmaili:2012vg}.
According to Ref.~\cite{Esmaili:2012vg},
in three years of data taking the sensitivity of the KATRIN experiment
will cover most of the region of the mixing parameters
of the sterile neutrino
allowed by the reactor and Gallium anomalies in the 3+1 framework.

\subsection{Neutrinoless double-$\beta$ decay}
\label{sub:bb}

The implications
of non-standard mainly sterile massive neutrinos at the eV scale
for neutrinoless double-$\beta$ decay experiments
have been studied by several authors
\cite{Goswami:2005ng,Goswami:2007kv,Barry:2011wb,Li:2011ss,Rodejohann:2012xd,Giunti:2012tn,Girardi:2013zra,Pascoli:2013fiz,Meroni:2014tba,Abada:2014nwa,Giunti:2015kza,Pas:2015eia}.

\begin{figure}[t]
\null
\hfill
\includegraphics*[width=0.49\linewidth]{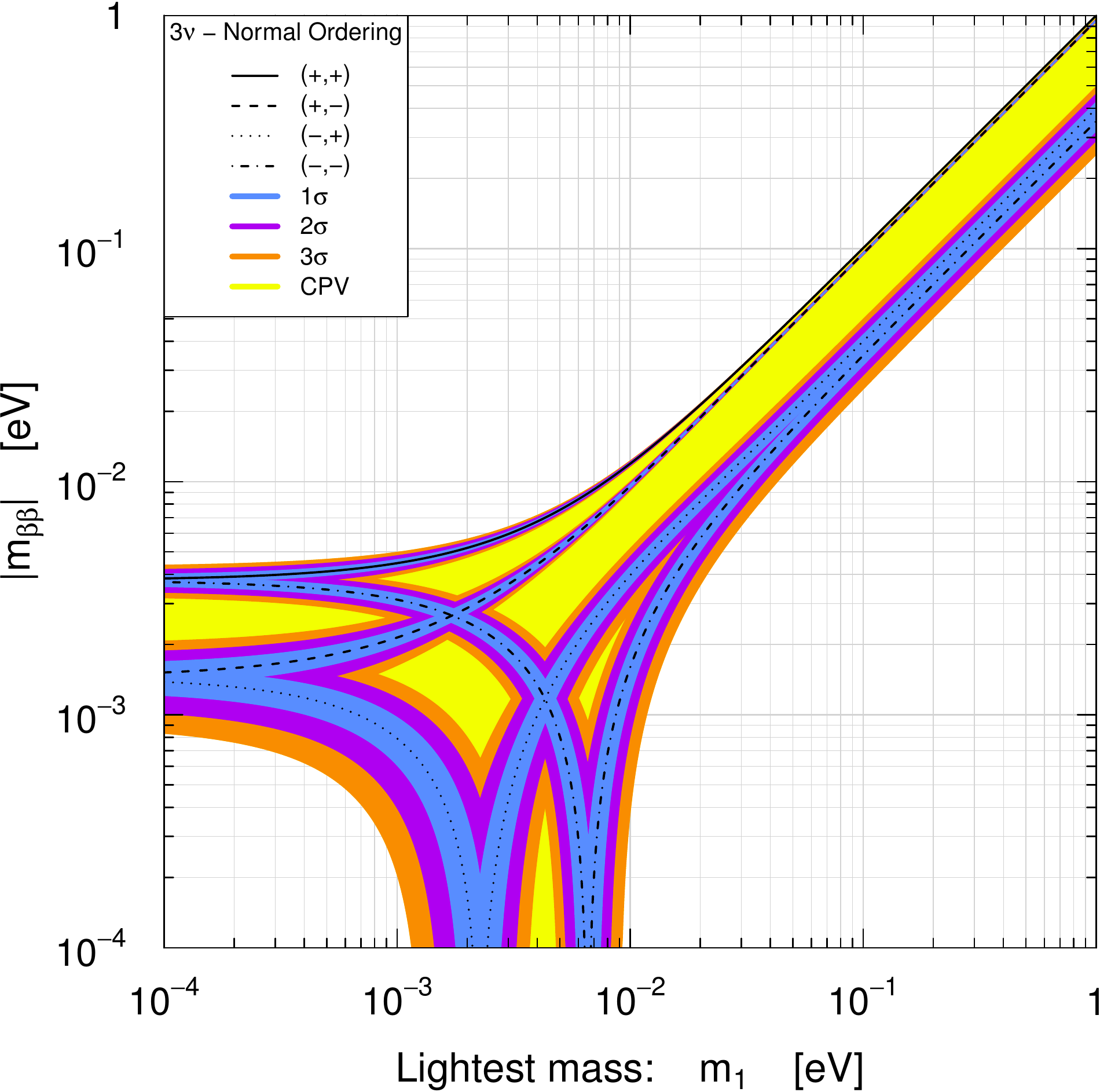}
\hfill
\includegraphics*[width=0.49\linewidth]{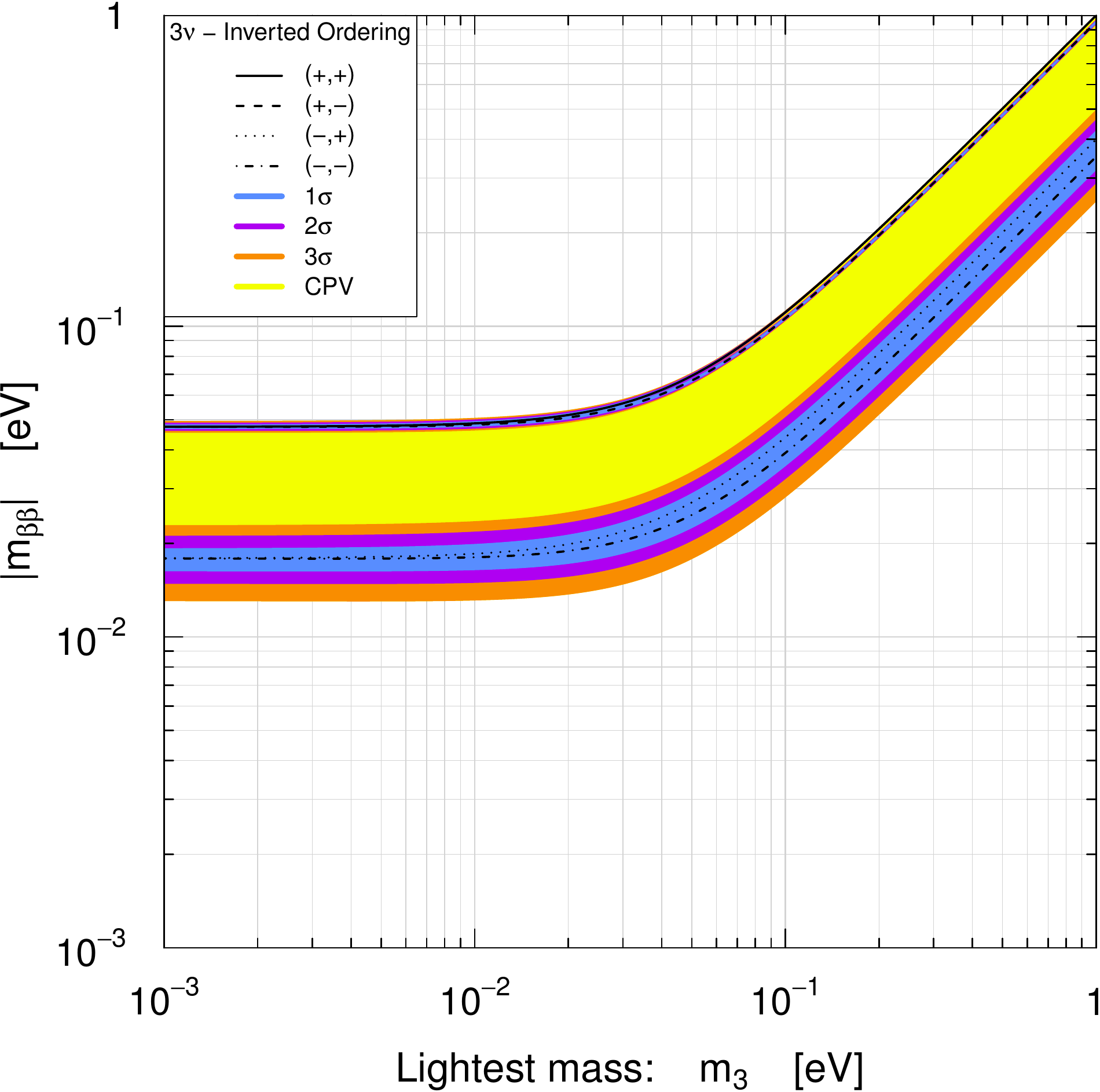}
\hfill
\null
\\
\null
\hfill
\includegraphics*[width=0.49\linewidth]{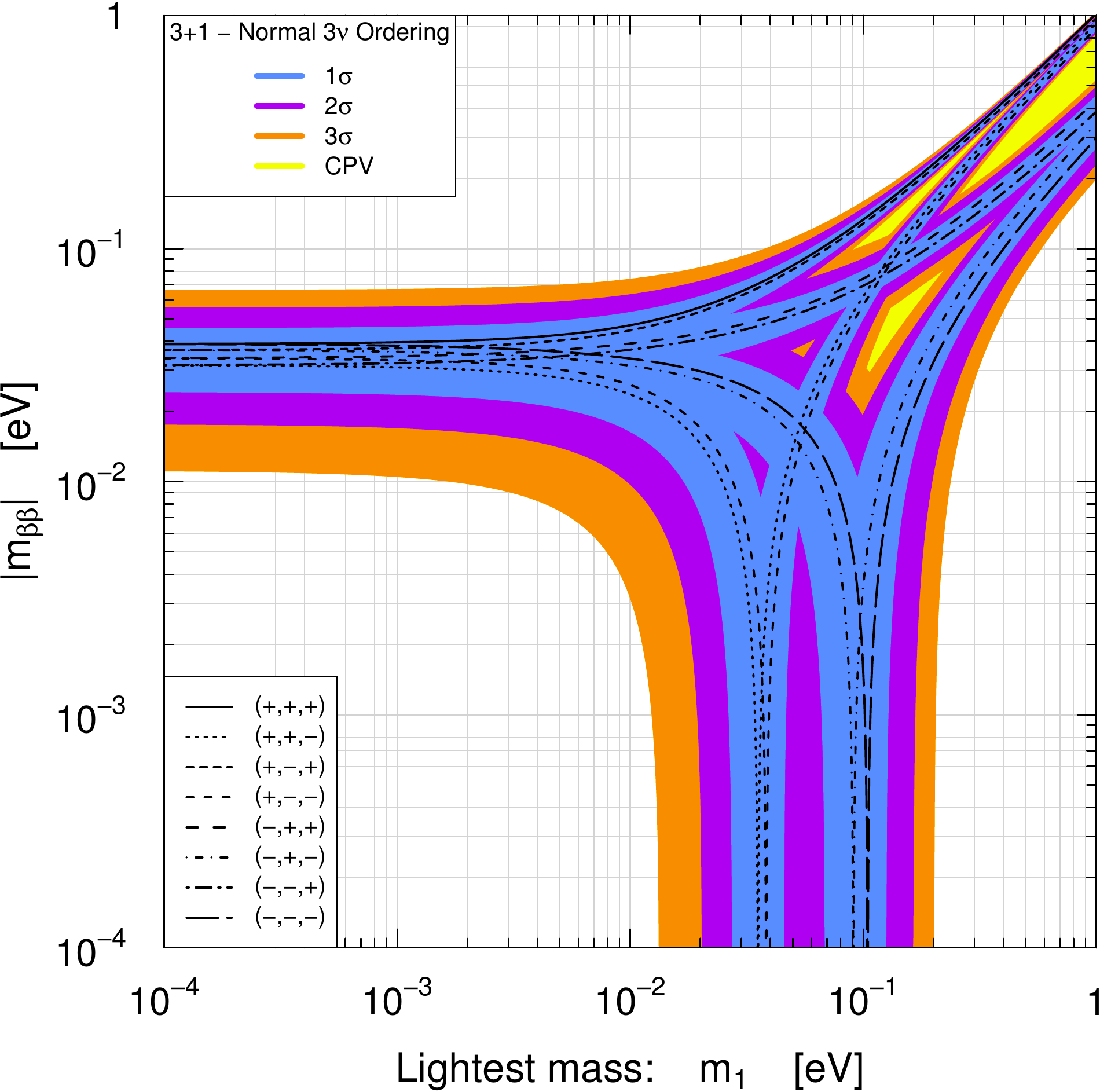}
\hfill
\includegraphics*[width=0.49\linewidth]{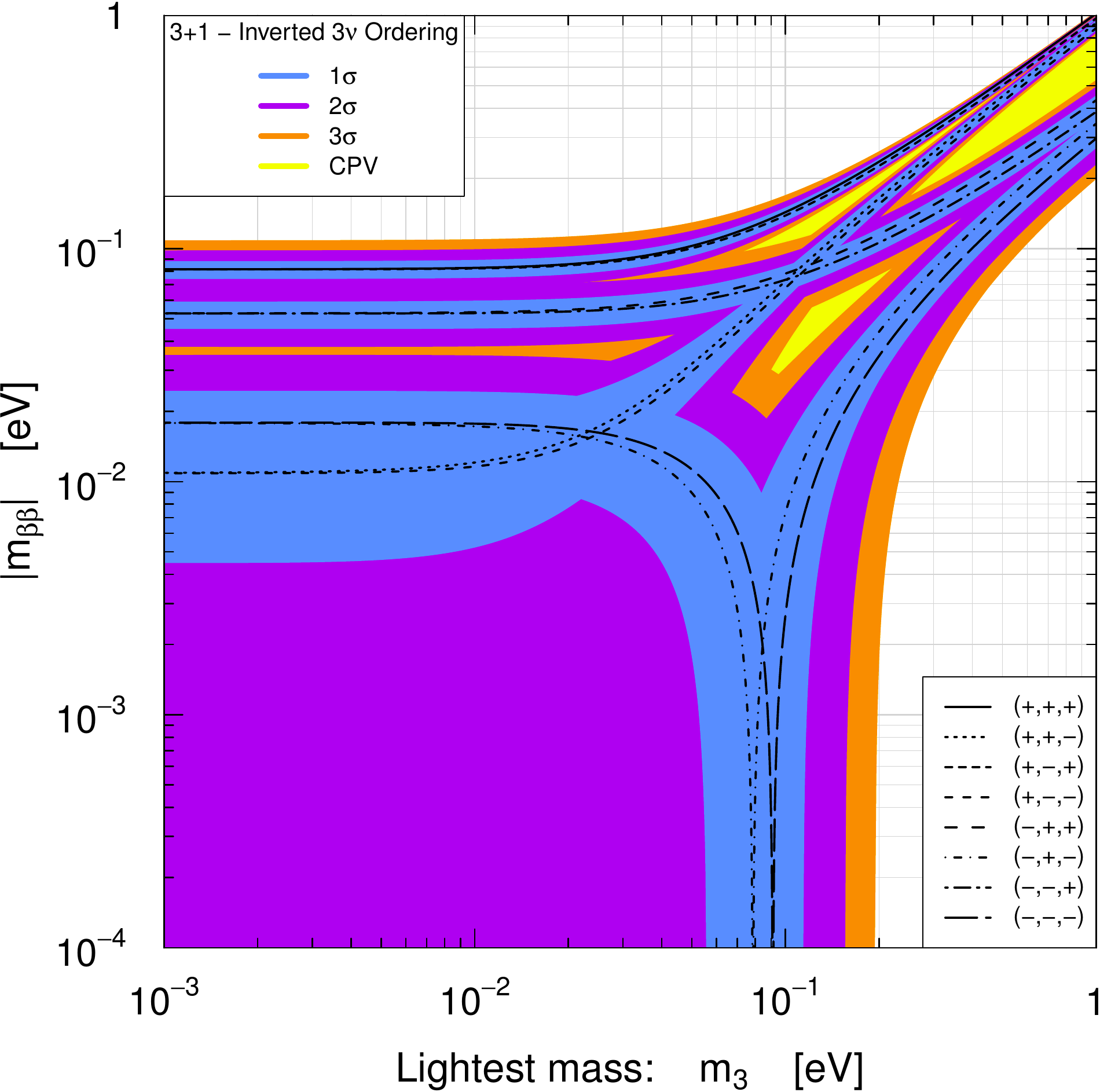}
\hfill
\null
\caption{ \label{fig:bb}
Value of the effective Majorana mass $|m_{\beta\beta}|$
as a function of the lightest neutrino mass
in the cases of $3\nu$ and 3+1 mixing with Normal and Inverted Ordering
of the three lightest neutrinos \cite{Giunti:2015kza}.
The signs in the legends indicate the signs of
$e^{i\alpha_{2}}, e^{i\alpha_{3}} , e^{i\alpha_{4}} = \pm1$
for the cases in which CP is conserved.
The intermediate yellow regions are allowed only in the case of CP violation.
}
\end{figure}

If massive neutrinos are Majorana particles
(see the recent reviews in Refs.~\cite{Bilenky:2014uka,Pas:2015eia}),
in the case of 3+1 mixing
the rate of neutrinoless double-$\beta$ decay
is proportional to the square of the effective Majorana mass
\begin{equation}
|m_{\beta\beta}|
=
\left|
|U_{e1}|^2 \, m_{1}
+
|U_{e2}|^2 \, e^{i\alpha_{2}} \, m_{2}
+
|U_{e3}|^2 \, e^{i\alpha_{3}} \, m_{3}
+
|U_{e4}|^2 \, e^{i\alpha_{4}} \, m_{4}
\right|
.
\label{mbb}
\end{equation}
In this expression there are three completely unknown complex phases
$\alpha_{2}$, $\alpha_{3}$, $\alpha_{4}$
which depend on the Majorana phases in the neutrino mixing matrix
(see, for example, Refs.~\cite{Giunti:2007ry,Bilenky:2010zza,Xing:2011zza}).
These unknown complex phases
can generate cancellations between the different mass contributions.
Figure~\ref{fig:bb}
shows the range of allowed values of $|m_{\beta\beta}|$
as a function of the lightest neutrino mass
in the cases of $3\nu$ and 3+1 mixing with Normal and Inverted Ordering
of the three lightest neutrinos \cite{Giunti:2015kza}.
The $3\nu$ mixing parameters are those in Tab.~\ref{tab:global3nu}
and the sterile neutrino mixing is that obtained in the pragmatic 3+1-PrGLO global fit
of short-baseline neutrino oscillation data
discussed in Section~\ref{sec:global}.

From Fig.\ref{fig:bb}
one can see that the presence of an additional
massive neutrinos at the eV scale
can change dramatically
the predictions for the possible range of values
of $|m_{\beta\beta}|$
\cite{Goswami:2005ng,Goswami:2007kv,Barry:2011wb,Li:2011ss,Rodejohann:2012xd,Giunti:2012tn,Girardi:2013zra,Pascoli:2013fiz,Meroni:2014tba,Abada:2014nwa,Giunti:2015kza,Pas:2015eia}.

\section{Cosmology}
\label{sec:cosmo}

In this Section we review the cosmological effects of a massive neutrino at the eV scale
in the standard \lcdm cosmological framework
(see also the recent reviews in Refs.~\cite{Archidiacono:2013fha,Lesgourgues:2014zoa}
and the book in Ref.~\cite{Lesgourgues-Mangano-Miele-Pastor-2013}).
Since such a massive neutrino is mostly sterile,
as explained in Section~\ref{sec:mixing},
its mass is traditionally denoted by the symbol $m_{s}$
in cosmological discussions.
In this section we use this notation,
keeping in mind that its real meaning in the 3+1 mixing scheme is
$m_{s}=m_{4}$.
Moreover,
in the discussion of the
combined analysis of cosmological data and short-baseline oscillation data
we consider
$m_{1},m_{2},m_{3} \ll m_{4}$,
so that
$m_{s} = m_{4} \simeq \sqrt{\Delta{m}^2_{41}} = \sqrt{\Delta{m}^2_{\text{SBL}}}$.

After introducing
in Subsection~\ref{sub:parameterization}
the parameterization of the neutrino density,
in Subsection~\ref{sub:radiationeffects}
we review the effects of light sterile neutrinos
on the observables generated in the first part of the evolution of the Universe,
when the sterile neutrinos were relativistic:
the Cosmic Microwave Background (CMB)
and
the nuclear abundances produced by Big Bang Nucleosynthesis (BBN).
In Subsection~\ref{sub:masseffects}
we discuss the effects of light sterile neutrinos
on the formation of Large Scale Structures (LSS),
which occurred after the sterile neutrinos became non-relativistic.
Finally,
in Subsection~\ref{sub:cmbbounds}
we review the current cosmological bounds on light sterile neutrinos.

\subsection{Neutrino parameterization}
\label{sub:parameterization}

It is convenient to parametrize the neutrino contribution to
the radiation content in the early Universe in terms of
an effective number of degrees of freedom \Neff,
such that the total energy density of relativistic species
$\rho_{r}$ is given by
\begin{equation}\label{eq:neff}
\rho_{r}
=
\left[1+\frac{7}{8}\left(\frac{4}{11}\right)^{4/3} \Neff\right] \rho_{\gamma}
=
\left[ 1 + 0.2271 \Neff \right] \rho_{\gamma}
,
\end{equation}
where $\rho_{\gamma}$ is the energy density of photons.
One unit of $\Neff$ corresponds to the contribution of one single family of active neutrinos
which were in equilibrium in the early Universe and
passed through an instantaneous decoupling at a temperature of about 1 MeV.
The factor $7/8$ is for a fermionic degree of freedom,
while the factor $T_\nu/T_{\gamma}=(4/11)^{4/3}$ comes from the fact that after neutrino decoupling there is a entropy transfer
from electrons to photons due to $e^\pm$ annihilations,
which enhances the photon temperature, but not the neutrino temperature.
Since in the real Universe
the decoupling of neutrinos was not instantaneous and the neutrinos were not completely decoupled when the $e^\pm$ annihilation occurred,
the effective number of active neutrino is slightly larger than three: $\Neff=3.046$~\cite{Mangano:2005cc}.
Assuming that all the additional contributions to the effective number of relativistic species come from sterile neutrinos,
their contribution to the total radiation energy density is quantified by
$\DNeff=\Neff-3.046$,
which is given by
\begin{equation}
\label{eq:dneff}
\DNeff
=
\frac{\rho_{s}^{\mathrm{rel}}}{\rho_\nu}
=
\left[\frac{7}{8}\frac{\pi^2}{15}{T_{\nu}}^4\right]^{-1}
\frac{1}{\pi^2}
\int dp \, p^3 f_{s}(p)
,
\end{equation}
where
$\rho_\nu$ is the energy density for one active neutrino species,
$\rho_{s}^{\mathrm{rel}}$ is the energy density of sterile neutrinos when relativistic,
$p$ is the neutrino momentum, $f_{s}(p)$ is the momentum distribution and
$T_{\nu} = (4/11)^{1/3} \, T_{\gamma}$.

After they become non-relativistic, neutrinos contribute to the matter energy density
of the Universe.
The sterile neutrino contribution can be parameterized
in terms of the dimensionless number
(see, for example, Ref.~\cite{Acero:2008rh})
\begin{equation}\label{eq:omegas}
\omega_{s}
=
\Omega_{s} h^2
=
\frac{\rho_{s}}{\rho_{c}} \, h^2
=
\frac{h^2}{\rho_c}
\frac{m_{s}}{\pi^2}
\int dp \, p^2 f_{s}(p)
,
\end{equation}
where
$\rho_{s}$ is the energy density of non-relativistic sterile neutrinos,
$\rho_c$ is the critical density and
$h$ is the reduced Hubble parameter.
Alternatively, $\omega_{s}$ can be converted in the effective neutrino mass \cite{Ade:2013zuv}
\begin{equation}\label{eq:meffs}
\meff{s} = 94.1 \, \omega_{s} \, \mathrm{eV}
.
\end{equation}

All the quantities that we introduced depend on the neutrino momentum distribution $f_{s}(p)$.
It is important that if the light sterile neutrinos decouple from the rest of the plasma when they are still relativistic,
$f_{s}(p)$ does not depend on $m_{s}$.
It depends only on the production mechanism.
The simplest example is one species of light sterile neutrinos that
are generated by active-sterile oscillations in the early Universe
\cite{Dolgov:2003sg,Cirelli:2004cz,Melchiorri:2008gq,Hannestad:2012ky,Mirizzi:2013kva,Hannestad:2015tea}
with the same temperature of active neutrinos.
In this case $\DNeff=1$
and
$\omega_{s} \simeq m_{s}/(94.1\,\mathrm{eV})$.

If the light sterile neutrino thermalizes at a temperature $T_{s}=\alpha T_\nu$,
its momentum distribution is given by the standard Fermi-Dirac distribution
\begin{equation}\label{eq:thermalnu}
f_{s}(p)=\frac{1}{e^{p/T_{s}}+1}.
\end{equation}
In this case,
called ``thermal scenario'' (TH),
from Eqs.~\eqref{eq:dneff} and \eqref{eq:omegas} we obtain
\begin{equation}\label{eq:THquantities}
\DNeff = \alpha^4
,
\qquad
\omega_{s} = \alpha^3\,\frac{m_{s}}{94.1\,\mathrm{eV}}
,
\qquad
\meff{s} = \alpha^3 m_{s} = \DNeff^{3/4} m_{s}
.
\end{equation}

For a non-thermal sterile neutrino generation, there are several possible mechanisms.
A popular one is the non-resonant production scenario, also called ``Dodelson-Widrow scenario'' (DW) \cite{Dodelson:1993je},
which is motivated by early active-sterile neutrino oscillations in the limit of zero lepton asymmetry and small mixing angle.
In this scenario the sterile neutrino has a momentum distribution
\begin{equation}\label{eq:DWnu}
f_{s}(p)=\frac{\beta}{e^{p/T_\nu}+1},
\end{equation}
where $\beta$ is a normalization factor, leading to
\begin{equation}\label{eq:DWquantities}
\DNeff=\beta
,
\qquad
\omega_{s} = \beta\,\frac{m_{s}}{94.1\,\mathrm{eV}}
,
\qquad
\meff{s} = \beta m_{s} = \DNeff m_{s}
.
\end{equation}
The DW and the TH models have an exact degeneracy, since they are related by
$\alpha=\beta^{1/4}$ and $m_{s}^{\mathrm{TH}}=m_{s}^{\mathrm{DW}}\beta^{1/4}$ \cite{Colombi:1995ze,Cuoco:2005qr}.

\subsection{Physical effects as radiation in the early universe}
\label{sub:radiationeffects}

As relativistic components, additional neutrino degrees of freedom change the time of matter-radiation equality,
whose redshift $\zeq$ is given by
\begin{equation}\label{eq:zeq}
1+\zeq
=
\frac{\rho_{m}}{\rho_{r}}
=
\frac{\omega_{m}}{\omega_{r}}
=
\frac{\omega_{m}}{\omega_{\gamma}}\frac{1}{1+0.2271\Neff},
\end{equation}
where $\rho_{m}$ and $\rho_{r}$ are the matter and radiation densities,
$\omega_{i}=\Omega_{i} h^2$,
with
$\Omega_{i}=\rho_{i}/\rho_{c}$,
and we used Eq.~(\ref{eq:neff}).

A shift in the matter-radiation equality affects
the position and the shape of the acoustic peaks of the CMB
(see Ref.~\cite{Archidiacono:2013fha}).
At photon decoupling (also called ``recombination'')
the extra radiation component enhances the expansion rate
$H=\dot{a}/a$,
where $a(t)$ is the scale factor in the
Friedmann-Robertson-Walker metric
(see, for example, Ref.~\cite{Lesgourgues-Mangano-Miele-Pastor-2013}).
This increase of $H$
generates a decrease of the comoving sound horizon $r_{\text{s}} \propto H^{-1}$
\cite{Hou:2011ec}
and a reduction of the angular scale of the acoustic peaks $\theta_{\text{s}}=r_{\text{s}}/D_A$, where $D_A$ is the angular diameter distance,
leading to a shift of the CMB peaks towards higher multipoles
(see Fig.~2(a) of Ref.~\cite{Archidiacono:2013fha}).
In addition,
if matter-radiation equality is delayed,
the amplitude of the first CMB peak at $\ell\simeq200$ is increased by the early Integrated Sachs Wolfe (ISW) effect,
since decoupling occurs when matter domination is at an earlier stage
and the subdominant radiation component causes a slow decrease of the gravitational potential
(see Figs.~2(a) and 2(b) of Ref.~\cite{Archidiacono:2013fha}).

These effects of additional relativistic neutrinos can be partially compensated if other cosmological parameters are simultaneously varied.
For example, if the total matter density
$\omega_{m}$ is also increased without altering the baryon density $\omega_b$,
so that the ratio between odd and even CMB peaks is not altered,
according to Eq.~(\ref{eq:zeq}) $\zeq$ can be kept fixed
and the two effects discussed above do not appear.
However,
one cannot obtain exactly the same CMB spectrum as in the standard case,
because additional relativistic neutrinos
increase the Silk damping effect at high multipoles
\cite{Bowen:2001in,Bashinsky:2003tk,Hou:2011ec}.

Silk damping is a diffusion damping of
the oscillations in the plasma
which occurs at high-multipoles because the decoupling of the baryon-photon interactions is not an instantaneous process.
During the time when the decoupling occurs, radiation free-streams
and the temperature fluctuations at scales
smaller than the radiation free-streaming scale are damped, because on such scales photons can move freely
from underdensities to overdensities and vice versa.
The damping depends on the ratio $r_{\text{d}} / r_{\text{s}}$,
where $r_{\text{d}} \propto H^{-1/2}$ is the photon diffusion length at recombination
\cite{Hou:2011ec}.
Since at fixed
$\zeq$
we have
$H^2 \propto \rho_{r} = \left( 1 + 0.2271 \Neff \right) \rho_{\gamma}$,
an increase of $\Neff$
corresponds to an increase of $H$
and an increase of $r_{\text{d}} / r_{\text{s}} \propto H^{1/2}$,
which enhances the Silk damping at high-multipoles
\cite{Hou:2011ec}.

Another important effect is related to BBN:
the number of relativistic degrees of freedom fixes the expansion rate during BBN, that in turn fixes
the abundances of light elements.
BBN can thus give strong constraints on $\Neff$
from the observations of the primordial abundances of light elements
\cite{Steigman:2012ve,Iocco:2008va,Jacques:2013xr,Fields:2014uja}.
According to Ref.~\cite{Mangano:2011ar}, BBN limits the effective number of additional relativistic species to
$\DNeff<1$ at 95\% C.L., regardless of the inclusion of CMB constraints on the baryon density $\Omega_b h^2$.
More recently, the authors of Ref.~\cite{Cyburt:2015mya} obtained
$\DNeff<0.2$ at 95\% C.L.
considering the BBN and CMB data.

\subsection{Physical effects as massive component}
\label{sub:masseffects}

Massive neutrinos have an effect in the early Universe due to their velocity.
Until they are relativistic they free-stream
over distances of the order of the Hubble radius
$H^{-1}$.
After they become non-relativistic their comoving free-streaming length diminishes together with their velocity.
As a consequence, there is a maximum comoving free-streaming length
at the time of the non-relativistic transition,
which corresponds today to the length
$\lambda_{\text{nr}} = 2 \pi / k_{\text{nr}}$,
with the wavenumber
$k_{\text{nr}}$ given by
\cite{Lesgourgues-Mangano-Miele-Pastor-2013}
\begin{equation}\label{eq:knr}
k_{\text{nr}}\simeq
0.0178
\,
\Omega_{m}^{1/2}
\left(\frac{T_{\nu}}{T_{s}}\right)^{1/2}
\left(\frac{m_{s}}{1\,\mathrm{eV}}\right)^{1/2} h \, \mathrm{Mpc}^{-1}
.
\end{equation}
The neutrino free-streaming
produces a suppression in the matter power spectrum
at scales smaller than $\lambda_{\text{nr}}$
which has two causes:
the absence of the contribution of neutrino perturbations
and
a suppression of the growth of Cold Dark Matter (CDM) perturbations
(see, for example, Refs.~\cite{Giunti:2007ry,Lesgourgues-Mangano-Miele-Pastor-2013,Lesgourgues:2014zoa}).

\subsection{Current bounds from cosmology}
\label{sub:cmbbounds}

The best measurements of the CMB anisotropies
come from the second data release of the Planck satellite experiment
(Planck 2015) \cite{Adam:2015rua, Ade:2015xua},
which improve the results of the first data release
(Planck 2013) \cite{Ade:2013sjv,Ade:2013zuv}.
Concerning neutrinos, in Ref.~\cite{Ade:2015xua} the Planck collaboration presented constraints on
the sum of the active neutrino masses $\mnu$,
on the effective number of relativistic species $\Neff$,
plus joint constraints on ($\mnu$, $\Neff$) and
on ($\meff{s}$, $\DNeff$).

In the analysis of the Planck collaboration \cite{Ade:2015xua}
the constraints on $\Neff$ alone have been obtained considering massless neutrinos in the \lcdm+$\Neff$ model.
However, it is interesting to present them in order to understand how the constraints on $\Neff$ change
when different experimental results are taken into account.
Considering the temperature (Planck TT) and the low-$\ell$ polarization (lowP) data, $\Neff=3.13\pm0.32$
\cite{Ade:2015xua},
which is consistent with the standard three-neutrino value $\Neff=3.046$.
The inclusion of BAO observations \cite{Anderson:2013zyy, Ross:2014qpa, Beutler:2011hx} tightens slightly the constraint to
$\Neff=3.15\pm0.23$,
leading to the upper bound $\DNeff<1$ at more than $3\sigma$
\cite{Ade:2015xua}.

Considering both $\meff{s}$ and $\DNeff$ as free parameters in the fit of cosmological data
in a \lcdm+$\Neff$+$\meff{s}$ model,
the Planck collaboration obtained \cite{Ade:2015xua}
\begin{equation}\label{eq:neffmeffPlTTlowPlensBAO}
\left.
 \begin{aligned}
\Neff &< 3.7 \\
\meff{s} &< 0.52\,\,\text{eV}
 \end{aligned}
\ \right\} \ \ \mbox{95\%, \text{Planck TT+lowP+lensing+BAO.}}
\end{equation}
In this case, the constraint on $\Neff$ is not significantly changed with respect to the \lcdm+$\Neff$ model,
and $\DNeff=1$ is still excluded at more than 3$\sigma$.
High values for $\DNeff$ are allowed preferably for sterile neutrinos
with small effective masses, corresponding to high values of
the rms amplitude
$\sigma_{8}$
of linear fluctuations today at a scale of $8 h^{-1} \, \text{Mpc}$.
On the contrary, for $\meff{s} \simeq 1 \, \text{eV}$ one obtains small values of $\sigma_{8}$,
because neutrino free-streaming suppresses
the amplitude of fluctuations at small scales.
In fact,
as we will discuss in the following,
a massive sterile neutrino can help solving
the current tension
between CMB measurements and astrophysical determinations of $\sigma_{8}$
(the values obtained for $\sigma_{8}$ from weak lensing
\cite{Heymans:2012gg,Kilbinger:2012qz,Kitching:2014dtq,Fu:2014loa,MacCrann:2014wfa},
cluster counts \cite{Ade:2013lmv, Ade:2015fva},
cluster mass functions \cite{Vikhlinin:2008ym, Burenin:2012uy}
and
redshift space distortions \cite{Beutler:2013yhm, Chuang:2013wga, Samushia:2013yga}
are lower than CMB estimates \cite{Ade:2015xua}).

\begin{figure}
\begin{center}
\begin{minipage}[r]{0.49\linewidth}
\begin{center}
\subfigure[]{\label{fig:meffneff1}
\includegraphics*[width=\linewidth]{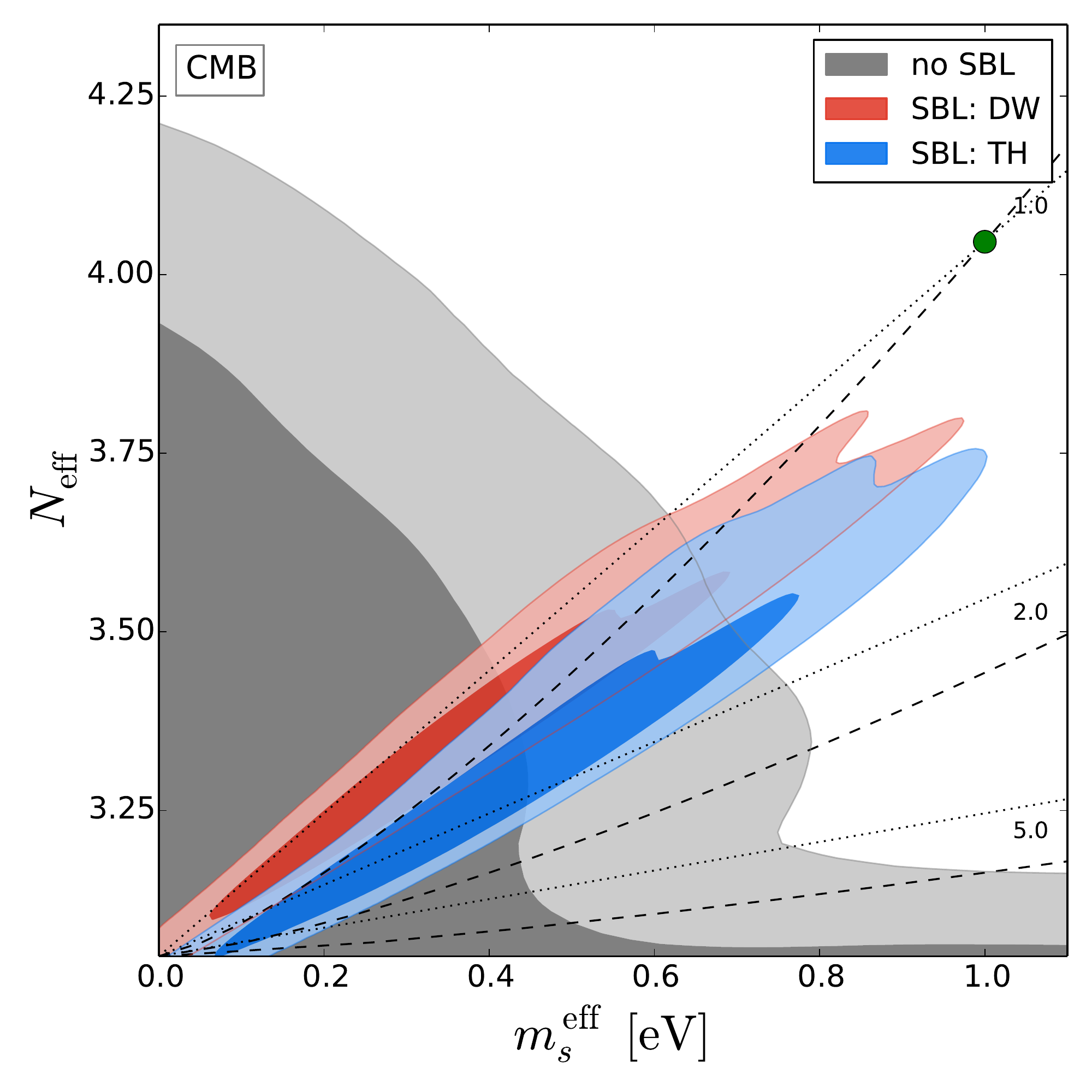}
}
\end{center}
\end{minipage}
\hfill
\begin{minipage}[l]{0.49\linewidth}
\begin{center}
\subfigure[]{\label{fig:meffneff2}
\includegraphics*[width=\linewidth]{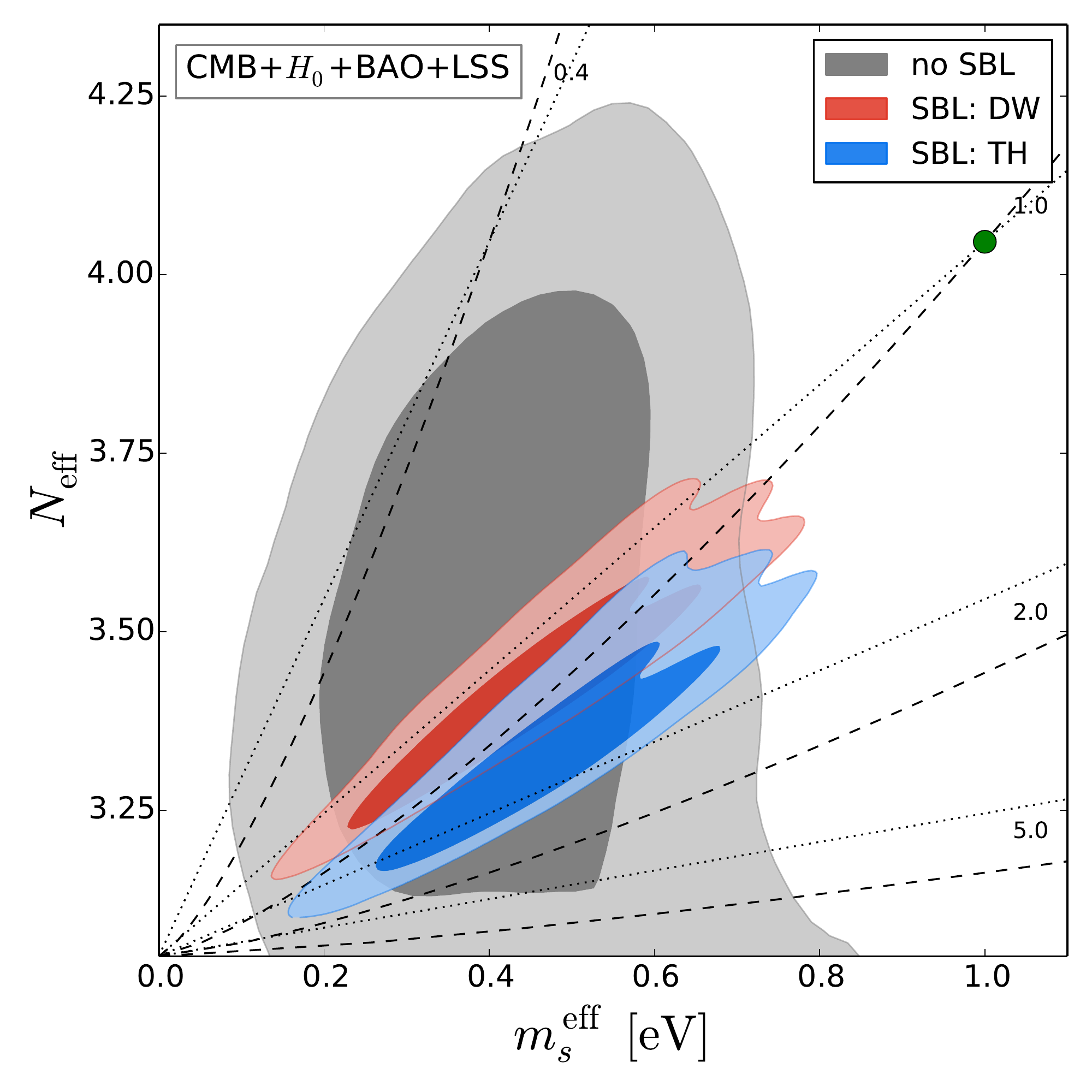}
}
\end{center}
\end{minipage}
\end{center}
\caption{ \label{fig:cosmo}
Allowed $1\sigma$ and $2\sigma$ regions in the $\meff{s}$-$\Neff$ plane for the \lcdm+$\Neff$+$\meff{s}$ model
with three different combinations for the SBL prior on $m_{s}$:
no SBL prior (gray),
with SBL prior in the DW scenario (red)
and in the TH scenario (blue).
The dotted (dashed) lines correspond to constant values, indicated in eV, for $m_{s}$ in the DW (TH) scenario.
The green circle represents a fully thermalized sterile neutrino with $m_{s}=1\,\text{eV}$.
The two figures,
adapted from Ref.~\cite{Gariazzo:2013gua},
have been obtained with different cosmological data sets.
\subref{fig:meffneff1}:
CMB data only (Planck 2013 TT at $\ell<2500$, WMAP $E$-mode polarization at $\ell<23$, ACT and SPT data at high $\ell$).
\subref{fig:meffneff2}:
CMB data,
the prior $H_0=73.8\pm2.4\Hou$ from HST \cite{Riess:2011yx},
the BAO measurements of the 6dFGRS \cite{Beutler:2011hx},
SDSS-II \cite{Padmanabhan:2012hf}, and
BOSS DR9 \cite{Anderson:2012sa} surveys, and
the cluster mass distribution at low and high redshift calculated
from the data of the Chandra Cluster Cosmology Project \cite{Vikhlinin:2008ym}
with the likelihood implementation of Ref.~\cite{Burenin:2012uy}.
}
\end{figure}

Remember that we are dealing with the effective mass
$\meff{s}$,
which is related to $m_{s}$
by Eqs.~\eqref{eq:omegas} and \eqref{eq:meffs}
in the general case,
by Eq.~\eqref{eq:THquantities} in the TH scenario,
and by Eq.~\eqref{eq:DWquantities} in the DW scenario.
Therefore,
in spite of the stringent upper bound (\ref{eq:neffmeffPlTTlowPlensBAO}) on
$\meff{s}$,
it is still possible to have a neutrino with $m_{s}\approx1$~eV
if $\DNeff$ is small.
However, if the sterile neutrinos
are generated by active-sterile oscillations in the early Universe
\cite{Dolgov:2003sg,Cirelli:2004cz,Melchiorri:2008gq,Hannestad:2012ky,Mirizzi:2013kva,Hannestad:2015tea},
they are fully thermalized well before CMB decoupling, resulting in $\DNeff\simeq1$,
which is disfavored by the bound
(\ref{eq:neffmeffPlTTlowPlensBAO})\footnote{
Previous bounds on $\DNeff$ for sterile neutrinos at the eV mass scale
have been discussed in Refs.~\cite{Dodelson:2005tp,Hamann:2010bk,Hamann:2011ge,Giusarma:2011zq,Joudaki:2012uk,Jacques:2013xr}.
}.
This problem led several authors to propose new mechanisms
that can relieve the tension:
a large lepton asymmetry
\cite{Foot:1995bm,Foot:1996qc,Bell:1998sr,Bell:1998ds,Shi:1999kg,DiBari:1999vg,DiBari:2001jk,Chu:2006ua,Hannestad:2012ky,Mirizzi:2012we,Saviano:2013ktj,Hannestad:2013pha},
new neutrino interactions \cite{Bento:2001xi,Dasgupta:2013zpn,Hannestad:2013ana,Bringmann:2013vra,Cherry:2014xra,Ko:2014bka,Archidiacono:2014nda,Saviano:2014esa,Mirizzi:2014ama,Tang:2014yla,Chu:2015ipa},
entropy production after neutrino decoupling \cite{Ho:2012br},
neutrino decay \cite{Gariazzo:2014pja},
very low reheating temperature \cite{Gelmini:2004ah,Smirnov:2006bu},
time varying dark energy components \cite{Giusarma:2011zq},
a larger cosmic expansion rate at the time of sterile neutrino production
\cite{Rehagen:2014vna},
inflationary freedom \cite{Gariazzo:2014dla}.

The authors of
Refs.~\cite{Archidiacono:2012ri,Archidiacono:2013xxa,Gariazzo:2013gua,Archidiacono:2014apa,Bergstrom:2014fqa}
performed combined fits
of cosmological data and short-baseline (SBL) oscillation data
in the framework of light sterile neutrino models.
The general result is that the sterile neutrino mass is constrained
at the scale of 1~eV by short-baseline oscillation data
and the full thermalization of the sterile neutrinos
in the early Universe is disfavored by cosmological data.
As an example, in Fig.~\ref{fig:meffneff1}, adapted from Ref.~\cite{Gariazzo:2013gua},
we show how the addition of the SBL prior on $m_{s}$
in the DW (red) and in the TH (blue) scenarios
changes the allowed regions in the $\meff{s}$-$\Neff$ plane
when the CMB data from
Planck 2013 (TT, $\ell<2500$),
WMAP ($E$-mode polarization at $\ell<23$) and
ACT/SPT (high-$\ell$ temperature spectrum)
are considered.
The dotted (dashed) lines correspond to constant values of $m_{s}$ in the DW (TH) scenario,
with the values in eV written near the lines.
One can note that since
CMB data force the allowed regions
to be as plotted in gray in Fig.~\ref{fig:meffneff1}
$\Neff$ and $\meff{s}$ cannot be both large at the same time.
Since the SBL prior requires $m_{s} \approx 1 \, \text{eV}$,
in the combined analysis
$\Neff$ is constrained below about 3.75 at $2\sigma$.

While CMB data disfavor eV-scale neutrino masses
and the presence of additional light sterile neutrino in cosmology,
LSS probes give some hints in favor of massive neutrinos.
Results in this direction were obtained, for example, in Ref.~\cite{Ade:2013lmv},
suggesting a preference for $\mnu=0.22\pm0.09$~eV for the active neutrinos
if the
Planck 2013 cluster counts through the Sunyaev-Zel'dovich (SZ) effect are considered together with CMB and BAO data.
Some of the analyses in the 2015 update of Planck SZ cluster counts
continue to indicate a preference for nonzero neutrino masses
(see, for example, Fig.~12 in Ref.~\cite{Ade:2015fva}).
These results arise from the fact that
the free-streaming of massive neutrinos can reduce the tension between the values of $\sigma_{8}$
obtained with CMB and LSS data
\cite{Heymans:2012gg,Kilbinger:2012qz,Kitching:2014dtq,Fu:2014loa,MacCrann:2014wfa,Ade:2013lmv,
Ade:2015fva,Vikhlinin:2008ym, Burenin:2012uy,Beutler:2013yhm, Chuang:2013wga, Samushia:2013yga}.
For the light sterile neutrinos, similar analyses were performed in the \lcdm+$\Neff$+$\meff{s}$ model
by several authors
\cite{Wyman:2013lza,Hamann:2013iba,Battye:2013xqa,Gariazzo:2013gua,Giusarma:2014zza,Archidiacono:2014apa,Bergstrom:2014fqa},
in spite of some tension between different datasets
\cite{Leistedt:2014sia}.

Figure~\ref{fig:meffneff2}, adapted from Ref.~\cite{Gariazzo:2013gua},
shows the allowed regions in the $\meff{s}$-$\Neff$ plane corresponding to those in Fig.~\ref{fig:meffneff1}
with the addition to the cosmological data set of
the prior on $H_0=73.8\pm2.4\Hou$ from HST \cite{Riess:2011yx},
the BAO measurements of the 6dFGRS \cite{Beutler:2011hx},
SDSS-II \cite{Padmanabhan:2012hf}, and
BOSS DR9 \cite{Anderson:2012sa} surveys, and
the cluster mass distribution at low and high redshift calculated
from the data of the Chandra Cluster Cosmology Project \cite{Vikhlinin:2008ym}
with the likelihood implementation of Ref.~\cite{Burenin:2012uy}.
Comparing the gray allowed regions in Figs.~\ref{fig:meffneff1} and \ref{fig:meffneff2}
one can see that the inclusion of LSS data favor values of
$\meff{s}$
around $0.2 - 0.6 \, \text{eV}$.
The inclusion of the SBL prior
(which restricts $m_{s}$ around 1 eV)
conserves this indication
by restricting the allowed value of
\Neff well below $\Neff=4$,
disfavoring a full thermalization of the sterile neutrino.

The preference for $\meff{s} \approx 0.2-0.6 \, \text{eV}$
is driven mainly by the inclusion
of the SZ cluster counts from Planck 2013,
but similar results are obtained with different datasets \cite{Battye:2013xqa,Gariazzo:2013gua}.
In all these cases, the presence of the light sterile neutrino has two effects:
its free-streaming reduces the fluctuations at small scales, and hence $\sigma_{8}$ is decreased,
while the presence of additional radiation in the primordial Universe leads to a slight increase of $H_0$.
The results obtained from the LSS measurements, however, can suffer from unaccounted systematics or some bias in the determination of the cluster masses.
Therefore,
a definite confirmation or rejection of the LSS preference for a massive sterile neutrino
must await future observations and analyses.
\begin{table}[t]
\begin{center}
\begin{tabular}{lccccc}
\hline
Project									& neutrino		& source		& $E$			& $L$		& status	\\
									& 			& 			& (MeV)			& (m)		&		\\
\hline
SAGE		\cite{Gavrin:2010qj,Gavrin:2015aca}			& $\nu_e$		& $^{51}$Cr		& $0.75$		& $\lesssim1$	& in preparation\\
LENS		\cite{Agarwalla:2010gd}					& $\nu_{e},\bar\nu_e$	& $^{51}$Cr, $^6$He	& $0.75$, $\lesssim3.5$	& $\lesssim3$	& abandoned	\\
CeLAND		\cite{Cribier:2011fv,Gando:2013zoa}			& $\bar\nu_e$		& $^{144}$Ce		& $1.8-3$		& $\lesssim6$	& abandoned	\\
Daya Bay	\cite{Dwyer:2011xs}					& $\bar\nu_e$		& $^{144}$Ce		& $1.8-3$		& $1.5-8$	& proposal	\\
LENA		\cite{Novikov:2011gp}					& $\nu_e$		& $^{51}$Cr, $^{37}$Ar	& $0.75$, $0.81$	& $\lesssim90$	& abandoned	\\
CeSOX		\cite{Borexino:2013xxa,Gaffiot:2014aka,Gaffiot-NOW2014}	& $\bar\nu_e$		& $^{144}$Ce		& $1.8-3$		& $5-12$	& in preparation\\
CrSOX		\cite{Borexino:2013xxa,Gaffiot-NOW2014}			& $\nu_e$		& $^{51}$Cr		& $0.75$		& $5-12$	& proposal	\\
JUNO		\cite{An:2015jdp}					& $\bar\nu_e$		& $^{144}$Ce		& $1.8-3$		& $\lesssim32$	& proposal	\\
\hline
\end{tabular}
\caption{Main features of proposed source experiments.}
\label{tab:source}
\end{center}
\end{table}

\section{Conclusions and perspectives}
\label{sec:conclusions}

The reactor, Gallium and LSND anomalies can be explained by neutrino oscillations
if the standard three-neutrino mixing paradigm
is extended with the addition of light sterile neutrinos
which can give us important information on the new physics beyond the Standard Model.

As discussed in Section~\ref{sec:global},
the global fits of neutrino oscillation data
in the framework of mixing schemes with one or more sterile neutrinos
suffer from a tension between the results of appearance and disappearance
short-baseline neutrino oscillation experiments.
Moreover,
as discussed in Section~\ref{sec:cosmo},
the cosmological data indicate a tension between the necessity to have a sterile neutrino mass
at the eV scale and the expected full thermalization
of the sterile neutrinos through active-sterile oscillations in the early Universe.
Hence,
the possible existence of light sterile neutrinos
at the eV scale
is a hot topic of current research and discussions.

The severe appearance-disappearance tension found in the global analyses of short-baseline neutrino oscillation data
\cite{Kopp:2013vaa,Giunti:2013aea}
can be alleviated adopting the ``pragmatic approach'' advocated in Ref.~\cite{Giunti:2013aea},
in which the anomalous MiniBooNE low-energy excess of $\nu_{e}$-like events is neglected.
The cause of this excess
is going to be investigated in the MicroBooNE experiment at Fermilab
\cite{MicroBooNE-2007,Katori:2014qta,Szelc:2015dga}.

There is an impressive program of many experimental projects
aiming at testing the short-baseline oscillations
due to sterile neutrinos
(see also the reviews in
Refs.~\cite{Lhuillier:2014mna,Katori:2014vka,Lasserre:2014ita,Caccianiga:2015ega,Lhuillier:2015fga,Spitz:2015gga}).
It is convenient to divide them in the following three categories
(see, however, also the proposals in
Refs.~\cite{Espinoza:2013dsa,Agarwalla:2009em,Bungau:2012ys}
out of these categories):

\begin{description}

\item[Source experiments.]
These experiments use radioactive sources
of $\nu_{e}$ or $\bar\nu_{e}$
placed near or inside a large detector.
Table~\ref{tab:source}
presents a list of the projects which have been proposed
(see also Ref.~\cite{Caccianiga:2015ega}).
To our knowledge, only the
SAGE \cite{Gavrin:2010qj,Gavrin:2015aca}
and
CeSOX \cite{Borexino:2013xxa,Gaffiot:2014aka,Gaffiot-NOW2014}
experiments
are in preparation.

\item[Reactor experiments.]
These experiments use a reactor $\bar\nu_{e}$ source with a detector placed at a distance of the order of 10 m.
There are several experiments in preparation,
as shown by the list in Tab.~\ref{tab:reactor}
(see also Ref.~\cite{Lhuillier:2015fga}).

\item[Accelerator experiments.]
These experiments,
listed in Tab.~\ref{tab:accelerator}
(see also Ref.~\cite{Spitz:2015gga}),
aim at checking the short-baseline
$\bar\nu_{\mu}\to\bar\nu_{e}$
LSND signal
(see Section~\ref{sub:LSND})
in both neutrino
($\nu_{\mu}\to\nu_{e}$)
and antineutrino
($\bar\nu_{\mu}\to\bar\nu_{e}$)
mode
(see also the nuSTORM proposal in Ref.~\cite{Adey:2014rfv}).
They will also look for the associated
$\nu_{\mu}$ and $\bar\nu_{\mu}$
disappearance
(see also the NESSiE proposal in Ref.~\cite{Stanco:2013dha}).

\end{description}

\begin{table}[t]
\begin{center}
\begin{tabular}{lcccc}
\hline
Project							& $P_{th}$	& $M_{target}$	& $L$		& Depth		\\
							& (MW)		& (tons)	& (m)		& (m.w.e.)	\\
\hline
Nucifer (FRA)		\cite{Cucoanes:2012jv}		& $70$		& $0.8$		& $7$		& $13$		\\
Stereo (FRA)		\cite{Abazajian:2012ys}		& $57$		& $1.75$	& $9-12$	& $18$		\\
Neutrino-4 (RUS)	\cite{Serebrov:2013yaa}		& $100$		& $1.5$		& $6-11$	& $10$		\\
Poseidon (RUS)		\cite{Derbin:2012kf}		& $100$		& $3$		& $5-8$		& $15$		\\
DANSS (RUS)		\cite{Danilov:2014vra}		& $3000$	& $0.9$		& $10-12$	& $50$		\\
SoLid (BEL)		\cite{Vacheret-AAP2013}		& $45-80$	& $3$		& $6-8$		& $10$		\\
Hanbit (KOR)		\cite{Yeo:2014spa}		& $2800$	& $1$		& $27$		& $10-23$	\\
Hanaro (KOR)		\cite{Yeo:2014spa}		& $30$		& $0.5$		& $6$		& few		\\
Prospect (USA)		\cite{Ashenfelter:2013oaa}	& $85$		& $1,10$	& $7,18$	& few		\\
CARR (CHN)		\cite{Guo:2013sea}		& $60$		& $\sim1$	& $7,11$	& few		\\
\hline
\end{tabular}
\caption{Main features of proposed reactor experiments.}
\label{tab:reactor}
\end{center}
\end{table}

The aim of most of these experiments is to reveal short-baseline oscillations in a robust way
by measuring distortions in the neutrino spectrum
or variations of the flavor neutrino detection probability as a function of distance.
In source experiments with monochromatic $\nu_{e}$'s generated by nuclear electron capture
(for example SAGE \cite{Gavrin:2010qj,Gavrin:2015aca} and CrSOX \cite{Borexino:2013xxa,Gaffiot-NOW2014}),
$\nu_{e}$ disappearance can be measured as a function of distance inside a sufficiently large detector.
In source experiments with a continuous $\bar\nu_{e}$ spectrum generated by nuclear $\beta$ decay
(for example CeSOX \cite{Borexino:2013xxa,Gaffiot:2014aka,Gaffiot-NOW2014})
and in reactor experiments
(for example Stereo \cite{Abazajian:2012ys})
both effects can be measured inside a sufficiently large detector with a sufficient energy resolution.
Some reactor experiments use two detectors at different distances
(for example Prospect \cite{Ashenfelter:2013oaa} and CARR \cite{Guo:2013sea})
or a movable detector
(for example DANSS \cite{Danilov:2014vra}).

For accelerator experiments
a crucial ingredient for reaching a robust result is the presence of ``near'' and ``far'' detectors
(as, for example, in the SBN \cite{Antonello:2015lea} experiment).
The near detector provides a normalization of the neutrino flux and cross section
which allows to measure the oscillations between the two detectors with small systematic uncertainty.

Since several of the experiments listed in Tabs.~\ref{tab:source}--\ref{tab:accelerator}
will be performed in the next years,
there are optimistic hopes that the important open problem of the existence of
light sterile neutrinos will be solved soon.
Let us emphasize that a positive result would be a major discovery
which would have a profound impact not only on neutrino physics,
but on our whole view of fundamental physics,
because sterile neutrinos are elementary particles beyond the Standard Model.
Hence,
their existence would prove that there is new physics beyond the Standard Model at low-energies
and their properties can give important information on this new physics.

For neutrino physics,
a discovery of the existence of light sterile neutrinos would open a rich field of experimental and theoretical research
on the properties of the sterile neutrinos, their mixing with the active neutrinos and their
role in
neutrino experiments
(e.g. in solar
\cite{Dooling:1999sg,Giunti:2000wt,Giunti:2009xz,Palazzo:2011rj,Palazzo:2012yf,Giunti:2012tn,Palazzo:2013me,Long:2013hwa,Long:2013ota,Kopp:2013vaa},
long-baseline
\cite{deGouvea:2014aoa,Klop:2014ima,Berryman:2015nua,Gandhi:2015xza,Palazzo:2015gja},
and atmospheric
\cite{Goswami:1995yq,Bilenky:1999ny,Maltoni:2002ni,Choubey:2007ji,Razzaque:2011ab,Razzaque:2012tp,Gandhi:2011jg,Esmaili:2012nz,Esmaili:2013cja,Esmaili:2013vza,Rajpoot:2013dha}
neutrino experiments)
in astrophysics
(e.g. in supernova neutrino experiments
\cite{Caldwell:1999zk,Peres:2000ic,Sorel:2001jn,Choubey:2006aq,Choubey:2007ga,Tamborra:2011is,Wu:2013gxa,Esmaili:2014gya,Warren:2014qza}
and indirect dark matter detection
\cite{Esmaili:2012ut}),
high-energy cosmic neutrinos
\cite{Cirelli:2004cz,Donini:2008xn,Barry:2010en},
and in cosmology (see Section~\ref{sec:cosmo}).
We think that this is a great opportunity which stimulates the great interest in the search for light sterile neutrinos.

\begin{table}[t]
\begin{center}
\begin{tabular}{lcccc}
\hline
Project										& $P$		& $M_{target}$	& $E$		& L		\\
										& (MW)		& (tons)	& (MeV)		& (m)		\\
\hline
SBN (USA)		\cite{Antonello:2015lea}				& $> 0.09$	& $112,89,476$	& $\sim 800$	& $110,470,600$	\\
J-PARC MLF (JPN)	\cite{Harada:2013yaa,Ajimura:2015yux}			& $\sim 1$	& $50$		& $\sim 40$	& $20$		\\
KPipe (JPN)		\cite{Axani:2015dha}					& $\sim 1$	& $684$		& $\sim 236$	& $32-152$	\\
nuPRISM (JPN)		\cite{Bhadra:2014oma}					& $\sim 1$	& $4000-8000$	& $200-1000$	& $1000-2000$	\\
IsoDAR-KamLAND (JPN)	\cite{Bungau:2012ys,Aberle:2013ssa,Conrad:2013ova}	& $0.6$		& $1000$	& $\sim 6.5$	& $10-40$	\\
IsoDAR-JUNO (CHN)	\cite{Conrad:2013ova,An:2015jdp}			& $0.6$		& $20000$	& $\sim 6.5$	& $20-100$	\\
OscSNS (USA)		\cite{Elnimr:2013wfa}					& $1.4$		& $450$		& $\sim 40$	& $50-70$	\\
\hline
\end{tabular}
\caption{Main features of proposed accelerator experiments.}
\label{tab:accelerator}
\end{center}
\end{table}

\section*{Acknowledgments}
The work of S. Gariazzo, C. Giunti and M. Laveder
is supported by the research grant {\sl Theoretical Astroparticle Physics} number 2012CPPYP7 under the program PRIN 2012 funded by the Ministero dell'Istruzione, Universit\`a e della Ricerca (MIUR).
The work of YFL is supported in part by the National Natural Science Foundation of China under Grant Nos. 11135009 and 11305193, by the Tianjin
Science Technology Research Funds of China under Grant No. 12JCYBJC3200, and by the joint project of RFBR (Russia, 15-52-53112) and NSFC (China, 11511130016).
E.M. Zavanin thanks the support of funding grants 2013/02518-7 and 2014/23980-3, S\~ao Paulo Research Foundation (FAPESP).


\end{document}